\documentclass[nobibnotes,nofootinbib,preprintnumbers,amsmath,amssymb,pre,floatfix,showkeys,superscriptaddress,twocolumn]{revtex4}
\usepackage{amsmath}
\usepackage{amssymb}
\usepackage[sort&compress]{natbib}
\usepackage{natbib}
\usepackage{enumerate}
\usepackage{graphicx}
\usepackage{dcolumn}
\usepackage{bm}
\usepackage{wasysym}
\usepackage{ae,aecompl}
\usepackage{color}

\usepackage[english]{babel}
\usepackage[latin2]{inputenc}
\usepackage[T1]{fontenc}

\newcommand{\PLT}{P_\mathrm{LT}}
\newcommand{\PFPT}{P_\mathrm{FPT}}
\newcommand{\PTTF}{P_\mathrm{TTF}}
\newcommand{\PTTC}{P_\mathrm{TTC}}
\newcommand{\LT}{\mathrm{LT}}
\newcommand{\FPT}{\mathrm{FPT}}
\newcommand{\TTF}{\mathrm{TTF}}
\newcommand{\TTC}{\mathrm{TTC}}

\newcommand{\addfig}[2]
{
\begin{figure}[ptb]
\centerline{\includegraphics[width=240pt]{#1}}
\caption{#2}
\label{fig:#1}
\end{figure}
}

\newcommand{\addtwofigs}[3]
{
\begin{figure*}[ptb]
\centerline{\includegraphics[width=220pt]{#1}\includegraphics[width=220pt]{#2}}
\caption{#3}
\label{fig:#1}
\end{figure*}
}

\newcommand{\addfourfigs}[5]
{
\begin{figure*}[ptb]
\centerline{\includegraphics[width=220pt]{#1}\includegraphics[width=220pt]{#2}}
\centerline{\includegraphics[width=220pt]{#3}\includegraphics[width=220pt]{#4}}
\caption{#5}
\label{fig:#1}
\end{figure*}
}

\def\ccc#1;#2{\left\langle #1 \left\vert #2 \right.\right\rangle}

\def\med #1{M\left[ #1 \right]}

\begin{document}

\preprint{}
\title{Diffusive behavior and the modeling of characteristic times in limit order executions}
\author{Zolt\'an Eisler}
\email{eisler@maxwell.phy.bme.hu}
\affiliation{Science \& Finance, Capital Fund Management, Paris, France}
\affiliation{Department of Theoretical Physics, Budapest University of Technology and Economics, Budapest, Hungary}
\author{J\'anos Kert\'esz}
\affiliation{Department of Theoretical Physics, Budapest University of Technology and Economics, Budapest, Hungary}
\affiliation{Laboratory of Computational Engineering, Helsinki University of Technology, Espoo, Finland} 
\author{Fabrizio Lillo}
\affiliation{Dipartimento di Fisica e Tecnologie Relative, Universit\`a di Palermo, Viale delle Scienze, I-90128, Palermo, Italy}
\affiliation{Santa Fe Institute, 1399 Hyde Park Road, Santa Fe, NM 87501, USA}
\affiliation{CNR-INFM, Unit\`a Operativa di Roma, Centro di Ricerca e Sviluppo SOFT, Roma, Italy}
\author{Rosario N. Mantegna}
\affiliation{Dipartimento di Fisica e Tecnologie Relative, Universit\`a di Palermo, Viale delle Scienze, I-90128, Palermo, Italy}
\affiliation{CNR-INFM, Unit\`a Operativa di Roma, Centro di Ricerca e Sviluppo SOFT, Roma, Italy}
\date{\today}
\keywords{Econophysics, Limit order book, First passage time, Brownian motion, Time to fill}

\begin{abstract} 
{We present an empirical study of the first passage time ($\FPT$) of order book prices needed to observe a prescribed price change $\Delta$, the time to fill ($\TTF$) for executed limit orders and the time to cancel ($\TTC$) for canceled ones in a double auction market. We find that the distribution of all three quantities decays asymptotically as a power law, but that of $\FPT$ has significantly fatter tails than that of $\TTF$. Thus a simple first passage time model cannot account for the observed $\TTF$ of limit orders. We propose that the origin of this difference is the presence of cancellations. We outline a simple model, which assumes that prices are characterized by the empirically observed distribution of the first passage time and orders are canceled randomly with lifetimes that are asymptotically power law distributed with an exponent $\lambda_\LT$. In spite of the simplifying assumptions of the model, the inclusion of cancellations is enough to account for the above observations and enables one to estimate characteristics of the cancellation strategies from empirical data.}
\end{abstract}

\maketitle

\section{Introduction}

Understanding the market microstructure is crucial for both theoretical and practical purposes \cite{biais.jfm2005}. On double auction markets the limit order book contains most of the information about the market microstructure and price discovery. Recently there has been considerable effort to investigate limit order book dynamics. Empirical studies \cite{biais.jf1995,handa.jf1996,harris.jfqa1996,kavajecz.jf1999,sandas.rfs2001,maslov.pa2001,challet.pa2001,lo.limitorder,potters.pa2003,hollifield.res2004, farmer.qf2004,zovko.farmer,mike.empirical2,weber.qf2005,hollifield.gains,ponzi.ph2006}
have been devoted to the search for the key determinants of price formation, the trading process and market organization.
A large number of papers have focused on modeling the limit order book with \cite{parlour.rfs1998, daniels.prl2003, foucault.flow, foucault.liquidity, rosu.dynamic} or without \cite{glosten.jf1004, chakravarty.jfi1995, seppi.rfs1997, luckock.qf2003} dynamics. 
Market microstructure studies consider a large number of aspects of the price discovery mechanism and these studies can greatly contribute to the success of the modeling of financial markets. The market mechanism, along with the complex interactions among market participants results in the emergence of a collective action of continuous price formation. Some of the studies have used an agent based modeling approach. Examples are market models described in terms of agents interacting through an order book based on simple rules  \cite{chiarella.iori,licalzi.qf2003} and models where the assumptions about the trading strategies are kept as minimal as possible \cite{daniels.prl2003,mike.empirical2}. One of the most striking findings was that even if trends and investor strategies are neglected, purely random trading may be adequate to describe certain basic properties of the order book \cite{zovko.farmer}. 

Most of the above papers focus on limit order executions, and very few deal with cancellations, even though the frequency of the two outcomes is comparable \cite{lo.limitorder}. The uncertainty of execution represents a primary source of risk \cite{chakrabarty.competing}. Another major risk factor is adverse selection, also known as "pick-off" risk. This risk is associated with the waiting time until order execution. During this period those with excess information can take advantage of the liquidity provided by the limit orders of less informed traders, and hence it is important to accurately quantify these waiting times. Lo et al. \cite{lo.limitorder} apply survival analysis to limit order data, and they find that the time between order placement and execution is very sensitive to the limit price, but not to the volume of the order. They also investigate the dependence on further explanatory variables such as the bid-ask spread and the volatility. The dynamics of the limit order book has also been investigated by using a joint model of executions and cancelations in a framework of competing risks\footnote{The notion of competing risks applies to problems where one deals with several "risks", i.e., random events, of which only the first one can be observed \cite{bedford.competing}. For example, limit orders are either executed or canceled and both events can be modeled by some random process. If an order gets canceled, one can no longer directly observe what time it would have been eventually executed, and vice versa. Thus it is not possible to independently estimate either process without a bias, if one simply ignores information from the other one.}. Within this approach Hollifield et al. \cite{hollifield.gains}, by using observations on order submissions and execution and cancellation histories, estimate both the distribution of traders' unobserved valuations for the stock and latent trader arrival rates. Chakrabarty et al. \cite{chakrabarty.competing} show that executions are more sensitive to price variation and less to volume variation than cancellations. This last work also analyzes the relationship between execution time and market depth.

In this paper we aim to go a step further, and combine the framework of competing risks with random walk theory. In particular, we analyze the difference observed between the time to fill a limit order, which is the time one had to wait before a limit order was executed, and the first passage time \cite{feller}, i.e., the time elapsed between an initial instant and the time when the transaction price crosses a given predefined threshold. In addition, the largest difference between our approach and most previous studies (e.g.,  Refs. \cite{lo.limitorder, chakrabarty.competing}) is that while those placed more emphasis on the typical values of execution and cancellation times, we will concentrate on the accurate description of the rare events, and the related asymptotic tail behavior of the distributions.

We observe that for a fixed price change the first passage time distributions of transaction price, best bid and best ask are quite well described asymptotically by the theoretical form expected for a Markov process with symmetric jump length distribution (including Brownian motion) \cite{feller, chechkin.JPA2003}.
The empirical time to fill of executed orders is smaller than the first passage time. We attribute this difference to canceled and expired orders. We propose a simple competing risks model, where limit orders are removed from the order book when either of two events happens: (i) when they are executed, this is modeled as the first time when the transaction price reaches the limit price, (ii) or when they are canceled, the time horizon of cancellations is modeled as a random process that is independent from price changes. In this framework we are able to predict constraints about the tail behavior of the time to fill and time to cancel probability densities. Our model also allows us to estimate the distribution of the time horizons of the placed limit orders. We show that the assumption of independence between the price changes and order cancellations, while it is a large simplification compared to real data, does not affect our conclusions significantly.

The paper is organized as follows. In Section~\ref{sec:book} we describe 
the investigated market and the variables of interest. In Section~\ref{sec:fpt} 
we study the first passage time and in Section~\ref{sec:ttfttc} the time to fill and the time to cancel. Section~\ref{sec:model} describes a simple limit order  
model and Section~\ref{sec:predictions} is devoted to testing the model empirically. Section~\ref{sec:moredelta} extends the result to limit orders placed inside the spread.  Section~\ref{sec:conclusions} discusses the validity of the assumptions and summarizes the results. {Finally, in the Appendices we show that the results are unchanged if time is measured in transactions.} Then we present a critical discussion of the fitting procedure we used to estimate the tail bahavior of the time to fill and time to cancel distributions.

\section{The dataset}
\label{sec:book}

The empirical analysis presented in this study is based on 
the trading data of the electronic market (SETS) of London Stock Exchange (LSE) 
during the year $2002$. These data can be purchased directly from the London Stock Exchange. We investigate $5$ highly liquid stocks, AstraZeneca (AZN), GlaxoSmithKline (GSK), Lloyds TSB Group (LLOY), Shell (SHEL), and Vodafone (VOD). Opening times of LSE are divided into three periods. The intervals 7:50--8:00 and 16:30--16:35 are called the opening and the closing auction, respectively. These follow different rules and thus also observe different 
statistical properties than the rest of the trading. Therefore we discarded 
limit orders placed during these times, and focused only on the periods of 
continuous double auction during 8:00--16:30. 
We also removed limit orders that were placed during 8:00--16:30 but were canceled (or expired) during the 
opening/closing auctions. We measure time intervals in trading time, i.e., we discard the 
time between the closing and the opening of the next day.\footnote{In our 
analyses, we removed the data of trading on September 20, 2002. This is because 
on that day very unusual trading patterns were observed, including an anomalous 
behavior of the bid-ask spread.} Finally, whenever we refer to prices we exclude all transactions that were executed on the SEAQ 
market\footnote{Many studies refer to this colloquially as the "upstairs" market.} and not in the limit order book.

We denote the best bid price\footnote{{In most of the literature the logarithm of the price is modeled, while throughout the paper we intentionally use price itself. Our study is concerned with very small price changes on the order of the spread,
when there is little difference between the two approaches. In our
case it is important to keep bare prices, as stocks have a finite tick
size (minimal price change). Taking bare prices enables us to classify
the orders into discrete categories by price difference. The size of
ticks depends on the stock, the possible values are $1/4$, $1/2$ or $1$ penny.}} by $b(t)$, the best ask price by $a(t)$ and the bid-ask spread is $s(t)=a(t)-b(t)$. Except for very special cases, there are already other limit orders waiting inside the book when one wants to place a new one. Let $b(t)-\Delta$ denote the price of a new buy limit order, and $a(t)+\Delta$ the price of a new sell limit order. Orders placed exactly at the existing best price correspond to $\Delta = 0$, orders placed inside the spread have $\Delta < 0$, while $\Delta > 0$ means orders placed "inside the book". It is possible to have so called crossing orders with such large negative values of $\Delta$ that they cross the spread, i.e., $\Delta < b(t)-a(t)$. These orders can be partially or fully executed immediately by limit orders from the other side of the book. Since a trader would place a crossing limit order to execute (at least part of) it immediately, we will not consider them as limit orders in our analysis.

Any limit order which was not executed can be canceled at any time by the trader who placed it. The order can also have a predetermined validity after which it is automatically removed from the book, this is called expiry. We will not distinguish between these mechanisms and we will call both of them cancellation.
Throughout the paper we will use ticks as units of price and all 
logarithms are $10$-base. 

\section{The first passage time}
\label{sec:fpt}

Let the latest transaction price of an asset 
at time $t_0=0$ be $S_0$. The first passage time \cite{feller} of price through a 
prescribed level $S_0+\Delta$ with some fixed $\Delta > 0$ is defined as 
the time $t$ of the first transaction when $S(t)\geq S_0+\Delta$. Similarly we can determine the first time after $t_0=0$ when the transaction price was below or equal to $S_0-\Delta$ and we will consider this time as another, independent observation of $t$. We will call the distribution of the quantity $t$ the first passage time distribution to a distance $\Delta$, and denote it by 
$P_{\mathrm{FPT};\Delta}(t)$.

Such first passage processes have been studied extensively \cite{redner}. For simplicity we will restrict ourselves to driftless processes. This is justified, because in real data for time horizons $t$ of up to a day the drift of the prices is negligible. This means that the ratio $|\mu|\sqrt{t}/\sigma$ is small (it is always less than $10^{-1}$ in our dataset), where $\mu$ is the mean price change over unit time, and $\sigma$ is the standard deviation of price changes during a unit time (i.e., the volatility). Throughout the paper we use real time\footnote{{We repeated the statistical analysis with transaction time and observed a similar power law decay of the first passage time for large times. The value of the power law exponent turns out to be different for real time analysis and transaction time analysis. See Appendix \ref{app:ttime} for details.}}.

For the following analysis of empirical data, it is useful to review the first passage time distribution for Brownian motion without drift. This is can be written as \cite{feller}
\begin{equation}P_{\mathrm{FPT};\Delta}(t) = \frac{\Delta}{\sqrt{2\pi\sigma^2}}t^{-3/2} 
\exp\left(-\frac{\Delta^2}{2\sigma^2t}\right),
\label{eq:fpt_brownian_full}
\end{equation}
which is the fully asymmetric $1/2$-stable distribution. 
 For any fixed $\Delta$ the asymptotics for long times is 
\begin{equation}	
P_{\mathrm{FPT};\Delta}(t)\propto t^{-3/2}.	
\label{eq:fpt_brownian}
\end{equation}

A recent study \cite{chechkin.JPA2003} has clarified that this asymptotic behavior is valid not only for Brownian motion but also for any Markov process with symmetric jump length distribution.\footnote{This result is consistent with the Sparre-Andersen theorem \cite{redner}. Alternative descriptions obtained for the asymptotic time dependence of the FPT of L\'evy flights which were  hypothesizing a dependence of the distribution exponent from the index of the L\'evy distribution have missed the fact that the method of images, which is extremely powerful in Gaussian diffusion, fails for L\'evy flight processes \cite{chechkin.JPA2003}. The behavior is of course more complex in the case of L\'evy random processes described by using a subordination scheme. In these cases the asymptotic behavior of first passage time depends on the complete properties of the subordination procedure \cite{sokolov}.} Of course, real price changes are not described by continuous values, and transactions and order submissions are also separated by finite waiting times, which a continuous time random walk formalism could take into account \cite{scalas.minireview, Montero}. However, in this paper we are interested in time intervals much longer than these waiting times, so the discrete aspects of the dynamics are negligible. Thus, we will model prices as if they varied continuously in time.

Let us now investigate empirically the first passage time behavior. The first 
passage time distribution for the transaction price, bid and ask when 
$\Delta = 1$ tick is shown in Fig. \ref{fig:Lfpt} for the stock GSK. The 
distribution is obtained by sampling the first passage time at each second. One 
can see that there are no significant differences in the behavior of the three 
prices. Qualitatively, the distribution is similar to Eq. \eqref{eq:fpt_brownian_full}, and the long time asymptotic of real data seems 
to decay approximately as $t^{-3/2}$. For times shorter than $1$ minute the 
curves significantly deviate both from the power law behavior and from the prediction of Eq. \eqref{eq:fpt_brownian_full}. We choose to fit the first passage time distribution with the function
\begin{equation}
P_{\mathrm{FPT};\Delta}(t) = \frac{Ct^{-\lambda_\mathrm{FPT}}}
{1+[t/T_\mathrm{FPT}(\Delta)]^{-\lambda_\mathrm{FPT}+\lambda'_\mathrm{FPT}}}.
\label{eq:pfptfit}
\end{equation}

This form, that we will use to fit also the other distributions introduced below, is characterized 
by two power law regimes. Normalization conditions of Eq. \eqref{eq:pfptfit} imply that  $\lambda_\mathrm{FPT}>1$ and $\lambda'_\mathrm{FPT}<1$. For $t\ll T_\mathrm{FPT}(\Delta)$ it is 
$P_{\mathrm{FPT};\Delta}(t)\propto t^{-\lambda'_\mathrm{FPT}}$, whereas for $t\gg 
T_\mathrm{FPT}(\Delta)$ it is $P_{\mathrm{FPT};\Delta}(t)\propto 
t^{-\lambda_\mathrm{FPT}}$.
We will discuss the motivations for choosing this form in Section IV and in the Appendix.

Table~\ref{tab:fpt} contains the fitted parameters 
$\lambda_\mathrm{FPT}$, $\lambda'_\mathrm{FPT}$, and $T_\mathrm{FPT}(\Delta)$ 
for $\Delta=1,\dots,4$ ticks. The difference between the actual values of $\lambda_\mathrm{FPT}$ and
$3/2$ from Eq. \eqref{eq:fpt_brownian} is small. Systematic deviations due to clustered volatility or the fluctuations of trading activity could not be identified. {For example, the asymptotic shape of the distribution does not change, even if time is measured in transactions instead of seconds (see Appendix \ref{app:ttime}).}

The observation that $\lambda_\mathrm{FPT} < 2$ implies that the theoretical mean and standard deviation of the first passage time distribution are infinite. Thus one should be careful with the interpretation of means calculated from finite samples. Throughout the paper we will rely on the determination of quantiles (e.g., the median) instead, which are always well-defined regardless of the shape of the distribution.
  
The inset of Fig.~\ref{fig:Lfpt} 
shows the median first passage time as a function of $\Delta$ for the five 
investigated stocks. The behavior is not exactly quadratic ($\Delta^2$) as one 
would expect from Eq.~\eqref{eq:fpt_brownian_full}. If prices followed a Brownian 
motion, the $q$-th quantile (${T_q}$) of the first passage time distribution 
would be
\begin{equation}	
T_q=\frac{\Delta^2}{2\sigma^2[\mathrm{erfc}^{-1}(q)]},	
\label{eq:fpttq}
\end{equation}
where the median ($\med{\FPT}$) corresponds to $q=0.5$. In reality, 
the power law behavior with $\Delta$ is less evident, as shown by the inset of Fig.~\ref{fig:Lfpt}. Assuming a behavior $\med{\FPT}\propto \Delta^{\eta}$
would require an exponent varying between $1.5$ and $1.8$ depending on the specific stock and the precise range of $\Delta$ used for the estimation of $\eta$. A similar deviation from the prediction of Brownian motion was reported in Ref. \cite{simonsen.optimal} in the analysis of closure index values sampled at a daily time horizon.

There are many differences between real prices and Brownian motion, and the above non-quadratic behavior can come from any of them: the non-Gaussian distribution of returns, the superdiffusivity of price, perhaps both or none. We have performed a series of shuffling experiments and preliminary results support the conclusion that the main role is played by the deviation from Gaussianity. This non-Gaussianity is well documented in the literature down to the scale of single transactions \cite{farmer.qf2004}. A similar effect was seen for Levy flights, whose increments are also very broadly distributed, and their value of $\eta$ can be different from $2$, and it is related to the index of the corresponding Levy distribution \cite{seshadri.pnas1982}.

\begin{table*}[tbp]
\centering
\begin{tabular}{c||ccc|ccc|ccc|ccc} \hline
stock & \multicolumn{3}{c|}{$\Delta = 1$} & \multicolumn{3}{c|}{$\Delta = 2$} & \multicolumn{3}{c|}{$\Delta = 3$} & 
\multicolumn{3}{c}{$\Delta = 4$} \\ \cline{2-13} & $\lambda$ & $\lambda'$ & $T$ 
& $\lambda$ & $\lambda'$ & $T$ & $\lambda$ & $\lambda'$ & $T$ & $\lambda$ & 
$\lambda'$ & $T$  \\ \hline
AZN & $1.50$ & $0.14$ & $58$ & $1.50$ & $0.22$ & $140$ & $1.50$ & $0.18$ & $240$ & $1.49$ & $0.11$ & $350$ \\
GSK & $1.52$ & $0.16$ & $62$ & $1.52$ & $0.18$ & $230$ & $1.50$ & $-0.02$ & $390$ & $1.48$ & $-0.21$ & $520$ \\
LLOY & $1.54$ & $0.22$ & $85$ & $1.55$ & $0.20$ & $280$ & $1.53$ & $0.01$ & $460$ & $1.51$ & $-0.12$ & $630$ \\
SHEL & $1.52$ & $0.20$ & $83$ & $1.53$ & $0.27$ & $160$ & $1.51$ & $0.02$ & $360$ & $1.51$ & $0.00$ & $450$ \\
VOD & $1.57$ & $0.43$ & $150$ & $1.54$ & $-0.19$ & $450$ & $1.49$ & $-0.69$ & $720$ & $1.51$ & $-0.66$ & $1500$ \\ 
\hline
\end{tabular}
\caption{Parameters of the fitting function \eqref{eq:pfptfit} for the 
distribution of first passage time for the five stocks. $\Delta$ is measured in 
ticks and all times are given in seconds. Typical standard errors for the quantities: 
$\pm 0.05$ for $\lambda_\FPT$, $\pm 0.05$ for $\lambda'_\FPT$, and $\pm 10\%$ for $T_\FPT$.}
\label{tab:fpt}
\end{table*}

\addfig{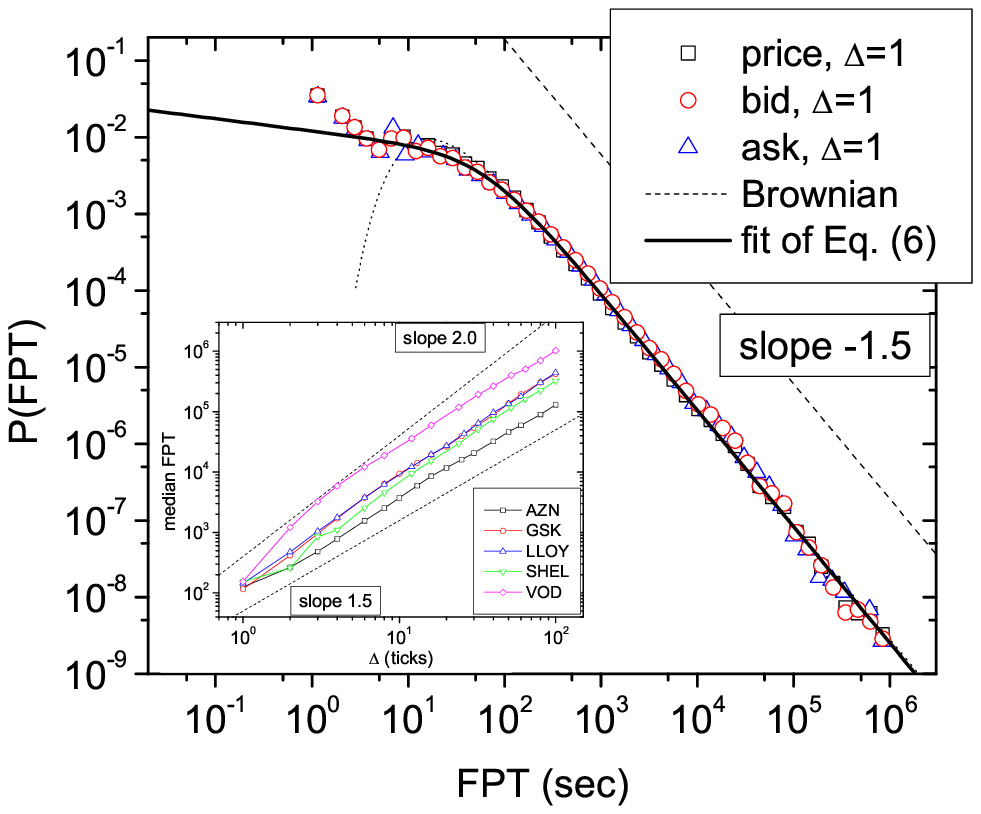}{First passage time distributions for the price, bid and ask quotes of GlaxoSmithKline (GSK), distance $\Delta = 1$ tick. The dotted line is the first passage time distribution for Brownian motion with volatility $\sigma=1/7$ penny$\times$ sec$^{-1/2}$. The thick solid line is a fit with Eq.~\eqref{eq:pfptfit} as given in Table \ref{tab:fpt}. The inset shows the median first passage time as a function of $\Delta$.}

\section{Time to fill, time to cancel}
\label{sec:ttfttc}

For an executed order the time elapsed between its placement and its complete execution is called \emph{time to fill}. Orders are often not executed in a single transaction, thus one can also define \emph{time to first fill}, which is the time from order placement to the first transaction this order participates in. Finally, 
for canceled orders one can define the \emph{time to cancel} which is the time 
between order placement and cancellation. The distribution of these three 
quantities will be in the following denoted by $ P_\mathrm{TTF}(t)$, 
$P_\mathrm{TTFF}(t)$, and $ P_\mathrm{TTC}(t)$, respectively.

\subsection{Properties of the distributions}

As a first characteristic of the order book, we investigate the 
distribution of time to fill and time to cancel for the stocks in our dataset. 
Fig. \ref{fig:GttfGSK} shows these distributions for GlaxoSmithKline 
(GSK) for different values of $\Delta$. Similarly to the first passage time, we 
fitted the empirical density with the function
\begin{equation}	
P_{\mathrm{TTF};\Delta}(t) = 
\frac{C't^{-\lambda_\mathrm{TTF}}}{1+[t/T_\mathrm{TTF}(\Delta)]^{-\lambda_\mathrm{TTF}+
\lambda'_\mathrm{TTF}}}.
\label{eq:pttffit}
\end{equation}

This form \eqref{eq:pttffit}, which we used to fit the FPT in the previous Section, is different from the more familiar generalized Gamma distribution used in Ref. \cite{lo.limitorder}. The reason for our choice is that we concentrate on the tail behavior of time distributions. According to our measurements the FPT, TTF and TTC distributions have fat tails, which can be well described by power laws. The generalized Gamma function has too slow convergence to a power law to describe the observed tails in the time range of our investigations. A detailed discussion of this problem is provided in Appendix \ref{app:fit}.

We also emphasize that in the present study we do not intend to discuss in detail the behavior on short time scales. We assume that this regime is simply characterized by the exponent $\lambda'$ only to perform a quick and efficient fit. This choice will have no direct relevance to our main conclusions, which always apply to the tails of the distribution. 

Nevertheless, in addition to the very good fit at large times the above formula gives for some cases an overall good description also at short times.
Table \ref{tab:ttf} shows the results for all five stocks. We find that 
$\lambda_\mathrm{TTF}$, which gives the asymptotic behavior of the distribution, ranges between $1.8$ and $2.2$ for up to $\Delta = 4$ ticks. This is greater than the value Ref. \cite{challet.pa2001} found for NASDAQ. The exponent $\lambda'_\mathrm{TTF}$ varies between $-0.4$ and $0.6$. Finally $T_\mathrm{TTF}$ typically grows with $\Delta$, as orders placed deeper into the book are executed later. We will return to this observation in Section \ref{sec:delta}. For $\Delta > 4$ the small number of limit orders in our sample does not allow us to make reliable  estimates for the shape of the distribution. Fig. 
\ref{fig:GttfGSK} also gives a comparison of four further stocks (AZN, LLOY, 
SHEL and VOD) to show that our findings are quite general. The distribution of time to first fill is indistinguishable from time to fill.

For time to cancel one finds a similarly robust behavior, also shown in Fig. 
\ref{fig:GttfGSK}. Its distribution is again well fitted by the form
\begin{equation}   	 
P_{\mathrm{TTC};\Delta}(t) = 
\frac{C''t^{-\lambda_\mathrm{TTC}}}{1+[t/T_\mathrm{TTC}(\Delta)]^{-\lambda_\mathrm{TTC}+
\lambda'_\mathrm{TTC}}},
\label{eq:pttcfit}
\end{equation}
where the long time asymptotics has an exponent $\lambda_\mathrm{TTC}$ ranging between $1.9$ and $2.4$. Unlike the case of $\lambda_\mathrm{TTF}$, the measured values of of $\lambda_\mathrm{TTC}$ are in agreement with those measured in Ref. \cite{challet.pa2001} for NASDAQ. All results concerning the time to cancel are given in Table \ref{tab:ttc}.

{As for the FPT, for both TTF and TTC the asymptotic power law behavior and the value of exponents is preserved if time is measured in transactions, see Appendix \ref{app:ttime}.}

\addfourfigs{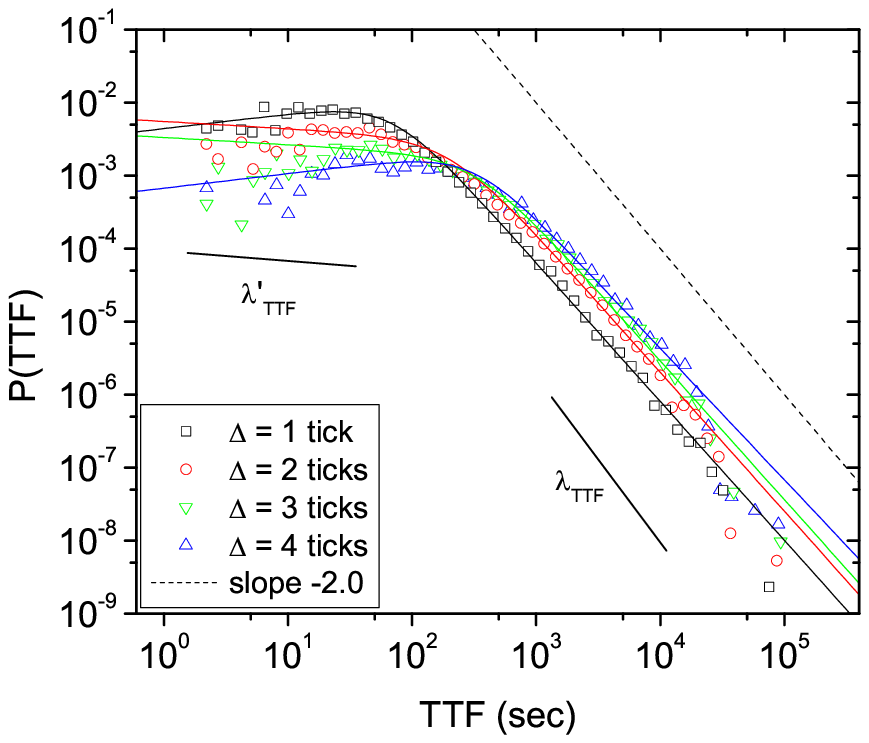}{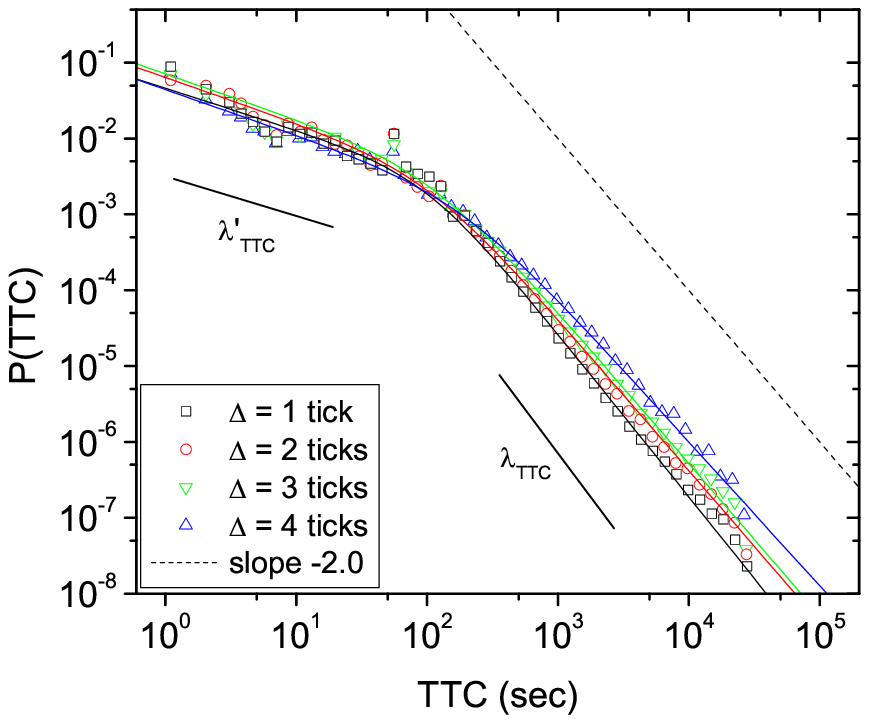}{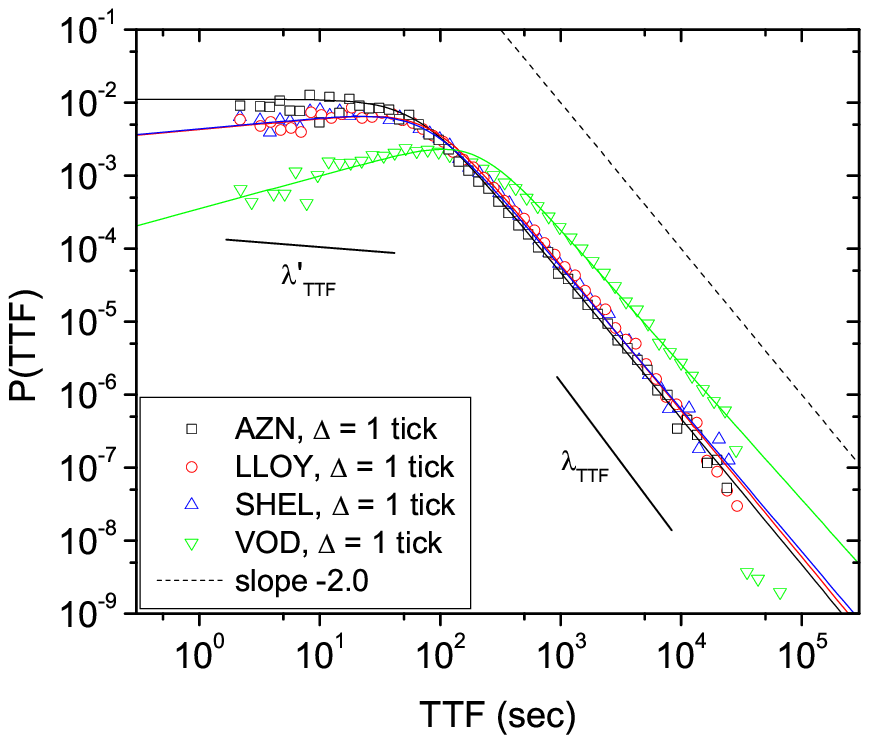}{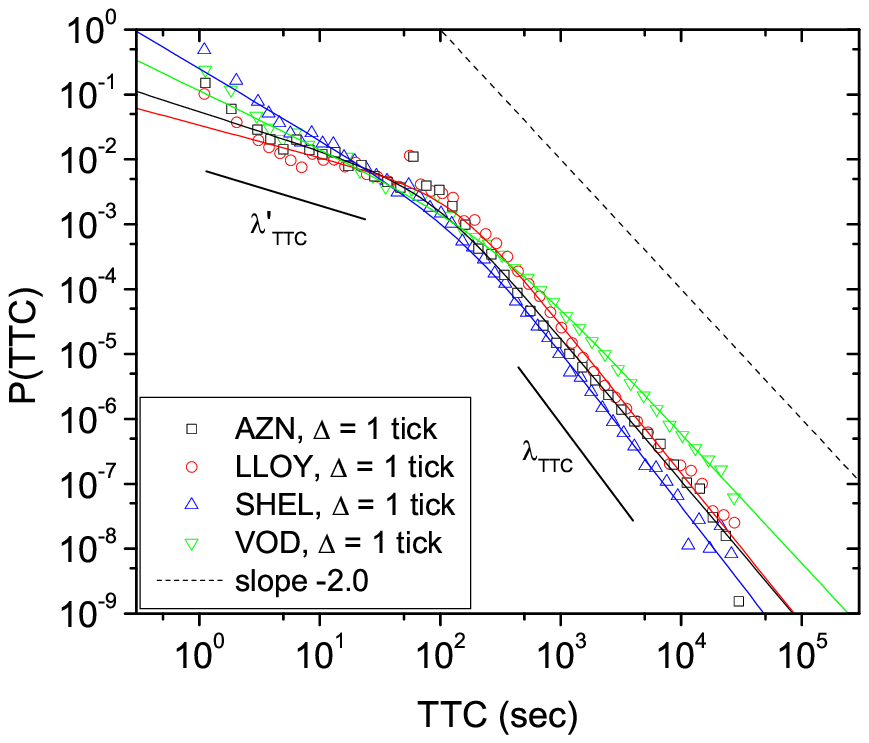}{\emph{Top left}: 
distribution of time to fill of GSK for $\Delta = 1\dots4$ ticks, 
and fits with Eq. \eqref{eq:pttffit}. The dashed line is a power law with 
exponent $-2.0$. \emph{Top right}: distribution of time to cancel of GSK for 
$\Delta = 1\dots4$ ticks, and fits with Eq. \eqref{eq:pttcfit}. The dashed line 
is a power law with exponent $-2.0$. \emph{Bottom left}: comparison of 
distributions of time to fill for three typical stocks, $\Delta = 1$ tick, and 
fits with Eq. \eqref{eq:pttffit}. The dashed line is a power law with exponent 
$-2.0$. \emph{Bottom right}: comparison of distributions of time to cancel for 
three typical stocks, $\Delta = 1$ tick, and fits with Eq. \eqref{eq:pttcfit}. 
The dashed line is a power law with exponent $-2.0$.}

\begin{table*}[tbp]\centering\begin{tabular}{c||ccc|ccc|ccc|ccc}\hline
stock & \multicolumn{3}{c|}{$\Delta = 1$} & \multicolumn{3}{c|}{$\Delta = 2$} & 
\multicolumn{3}{c|}{$\Delta = 3$} & \multicolumn{3}{c}{$\Delta = 4$} \\ 
\cline{2-13} & $\lambda$ & $\lambda'$ & $T$ & $\lambda$ & $\lambda'$ & $T$ & 
$\lambda$ & $\lambda'$ & $T$ & $\lambda$ & $\lambda'$ & $T$  \\ \hline
AZN & $2.0$ & $-0.0$ & $65$ & $1.9$ & $0.0$ & $100$ & $1.8$ & $-0.0$ & $120$ & $1.9$ & $0.0$ & $200$ \\
GSK & $1.9$ & $-0.2$ & $68$ & $1.9$ & $-0.2$ & $150$ & $1.8$ & $-0.4$ & $190$ & $1.8$ & $-0.3$ & $320$ \\
LLOY & $2.0$ & $-0.1$ & $85$ & $1.9$ & $-0.1$ & $160$ & $1.9$ & $-0.2$ & $240$ & $1.9$ & $-0.2$ & $350$ \\
SHEL & $1.9$ & $-0.1$ & $77$ & $1.9$ & $-0.2$ & $110$ & $1.9$ & $0.0$ & $270$ & $1.8$ & $-0.1$ & $250$ \\
VOD & $1.8$ & $-0.4$ & $190$ & $1.8$ & $-0.5$ & $490$ & $1.8$ & $-0.4$ & $980$ & -- & -- & -- \\ \hline
\end{tabular}	
\caption{Parameters of the fitting function \eqref{eq:pttffit} for the distribution of time to fill for the 
five stocks and $\Delta > 0$ ticks. All times are given in seconds. Data are 
missing where the statistics was inadequate for fitting. Typical standard errors for the quantities: $\pm 0.1$ for $\lambda_\TTF$, $\pm 0.1$ for 
$\lambda'_\TTF$, and $\pm 10\%$ for $T_\TTF$.}	
\label{tab:ttf}
\end{table*}

\begin{table*}[tbp]\centering\begin{tabular}{c||ccc|ccc|ccc|ccc}\hline
stock & \multicolumn{3}{c|}{$\Delta = 1$} & \multicolumn{3}{c|}{$\Delta = 2$} & 
\multicolumn{3}{c|}{$\Delta = 3$} & \multicolumn{3}{c}{$\Delta = 4$} \\ 
\cline{2-13} & $\lambda$ & $\lambda'$ & $T$ & $\lambda$ & $\lambda'$ & $T$ & 
$\lambda$ & $\lambda'$ & $T$ & $\lambda$ & $\lambda'$ & $T$  \\ \hline
AZN & $2.2$ & $0.6$ & $87$ & $2.2$ & $0.6$ & $90$ & $2.2$ & $0.6$ & $85$ & $2.2$ & $0.7$ & $100$ \\
GSK & $2.2$ & $0.5$ & $110$ & $2.0$ & $0.5$ & $90$ & $1.9$ & $0.5$ & $94$ & $1.9$ & $0.6$ & $170$ \\
LLOY & $2.3$ & $0.5$ & $130$ & $2.2$ & $0.4$ & $140$ & $2.0$ & $0.4$ & $120$ & $2.0$ & $0.5$ & $250$ \\
SHEL & $2.4$ & $1.1$ & $150$ & $2.3$ & $1.1$ & $140$ & $2.3$ & $1.1$ & $68$ & $2.2$ & $1.0$ & $56$ \\
VOD & $2.0$ & $0.9$ & $300$ & $2.1$ & $0.8$ & $1000$ & $2.2$ & $0.6$ & $1500$ & $1.9$ & $0.5$ & $1000$ \\ \hline
\end{tabular}	
\caption{Parameters of the fitting function \eqref{eq:pttcfit} for the distribution of time to cancel for the five stocks and $\Delta > 0$ ticks. All times are given in seconds. Data are missing when 
there were no orders at all, or the statistics was inadequate for fitting. 
Typical standard errors for the quantities: $\pm 0.1$ for $\lambda_\TTC$, 
$\pm 0.1$ for $\lambda'_\TTC$, and $\pm 25\%$ for $T_\TTC$.}	
\label{tab:ttc}
\end{table*}

\subsection{Comparison of characteristic times}

The emphasis of this paper is on the interplay between order execution, order 
cancellation and the first passage properties of price. To understand this 
relationship, consider the following argument proposed in Ref. 
\cite{lo.limitorder}. Imagine that there are no cancellations. Let a 
buy order be placed at the price $b_0-\Delta$, when the current best bid is at $b_0$ (the argument goes similarly 
for sell orders).
How much time does it take until this order is executed? It is certain 
that the order cannot be executed before the best bid decreases to $b_0-\Delta$, 
because until then there will always be more favorable offers in the book. On 
the other hand, once the price decreases to $b_0-\Delta-\epsilon$ where 
$\epsilon$ is the tick size of the stock, it is certain, that all possible 
offers at the price $b_0-\Delta$ have been exhausted, including ours. Therefore 
both time to fill and time to first fill for any order  placed at a distance $\Delta$ from the best offer is 
greater than the first passage time of price to a distance $\Delta$, and less 
than that to $\Delta+\epsilon$. Since this is true for every individual order, 
one expects the following inequality for the distribution functions of 
characteristic times:
\begin{eqnarray}	
\int_0^t  
P_{\mathrm{FPT};\Delta}(t')dt' \geq \int_0^t  P_{\mathrm{TTF};\Delta}(t')dt' 
\geq \nonumber \\ \int_0^t  P_{\mathrm{TTFF};\Delta}(t')dt' \geq \int_0^t  
P_{\mathrm{FPT};\Delta+\epsilon}(t')dt'.
\end{eqnarray}
Using the empirical distributions above, a straightforward calculation yields
\begin{equation}	
\lambda_\mathrm{FPT}=\lambda_\mathrm{TTFF}=\lambda_\mathrm{TTF},
\label{eq:lambdawrong}
\end{equation} 
which is in clear disagreement with the 
data, where pronouncedly $\lambda_\mathrm{FPT}<\lambda_\mathrm{TTF}\approx 
\lambda_\mathrm{TTFF}$. 
This inequality for the tail exponents means that one finds less 
orders with very long time to (first) fill than expected. The resolution of this 
apparent contradiction is that cancellations have to be taken into account: Orders which would have to wait too long before being executed are often canceled and thus removed 
from the statistic. The measurement of the cancellation time distribution suffers from the same bias. The observed distribution of time to cancel does not characterize how traders would
actually cancel their orders, because here the executed orders are missing from the statistics.

In Section \ref{sec:model} we will present a simple model that gives insight into the features pointed out so far. However, before doing so, we would like to present one further point concerning the empirical data.

\subsection{The role of entry depth}
\label{sec:delta}

How do order execution times change as a function of the entry depth $\Delta$? Similarly to first passage times, the empirical distributions found for time to fill/cancel have a slowly decaying tail such that the means might diverge. Therefore, in the following we will use the medians of all quantities as a measure of their typical value.

In Fig. \ref{fig:GlifedistGSK} we show that the median of time to fill is empirically well described by 
\begin{equation}
\med{\TTF} \propto \Delta^{1.4},
\end{equation}
which is quite different from the $\med{\TTF} \propto \Delta^2$ expected naively from Eq. 
\eqref{eq:fpttq} and a Brownian motion assumption, and also from the $\Delta^{1.5-1.8}$ behavior observed for the first passage time. We show that cancellations play an important role in these discrepancies.

Let us make a surrogate experiment with the data of the stock GSK. We select all filled orders, and from the time of 
their placement we calculate the {\it first} time when the transaction price becomes 
equal to or better than the price of the order. If one plots the median of this 
quantity versus the $\Delta$ of the orders, the resulting curve is 
indistinguishable from the median of time to fill [Fig. \ref{fig:GlifedistGSK}(left), curve labeled as "TTF/FPT filled ord"]. Thus the exponent $1.4$ does not come from the difference between order executions and first passage times.

In another surrogate experiment we keep the time of order placements, but shuffle the $\Delta$ values between orders. This way we destroy correlations between volatility and order placement. We record the corresponding first passage times. The resulting curve is labeled as "FPT shuff. all ord". This new curve now agrees with the first passage time of price [curve "FPT, price (book only)"] when $\Delta > 8$ ticks, which corresponds to a median time of about $1-2$ hours. The origin of the anomalous $\Delta$-dependence is, at least in the large $\Delta$ case, therefore the presence of cancellations. The explanation of other contributions requires more involved arguments which are beyond the scope of this paper.

The dependence of median time to cancel on the entry depth $\Delta$ has a less clear functional form, as shown by Fig. \ref{fig:GcancdistCOMPARISON}. While $\med{\TTC}$ appears to be a monotonically increasing function of $\Delta$, the curves for the different stocks show only a qualitative similarity. One of the reasons may be that different cancellation mechanisms are treated together. 

\addtwofigs{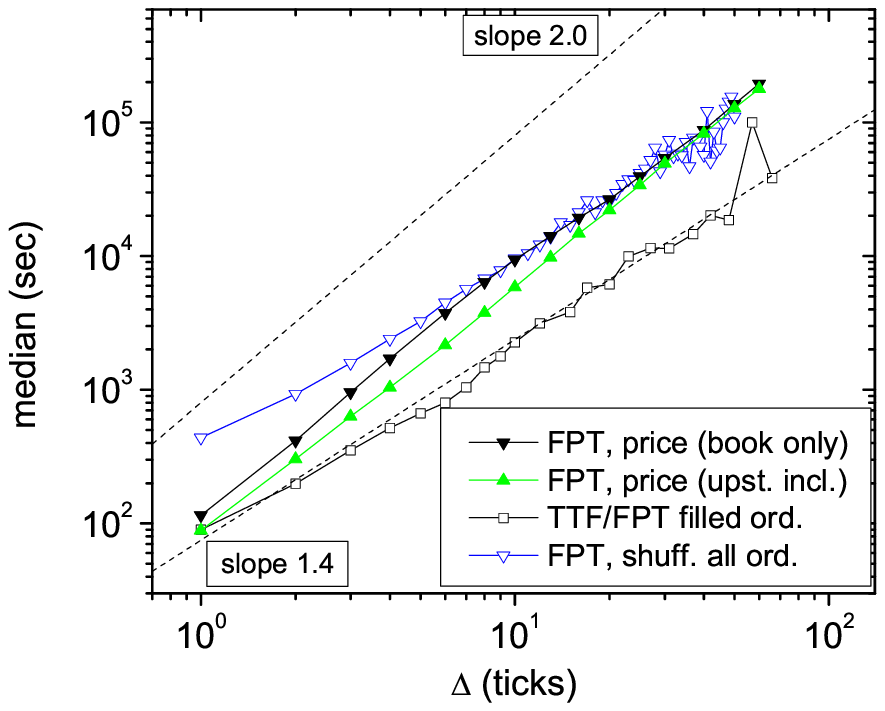}{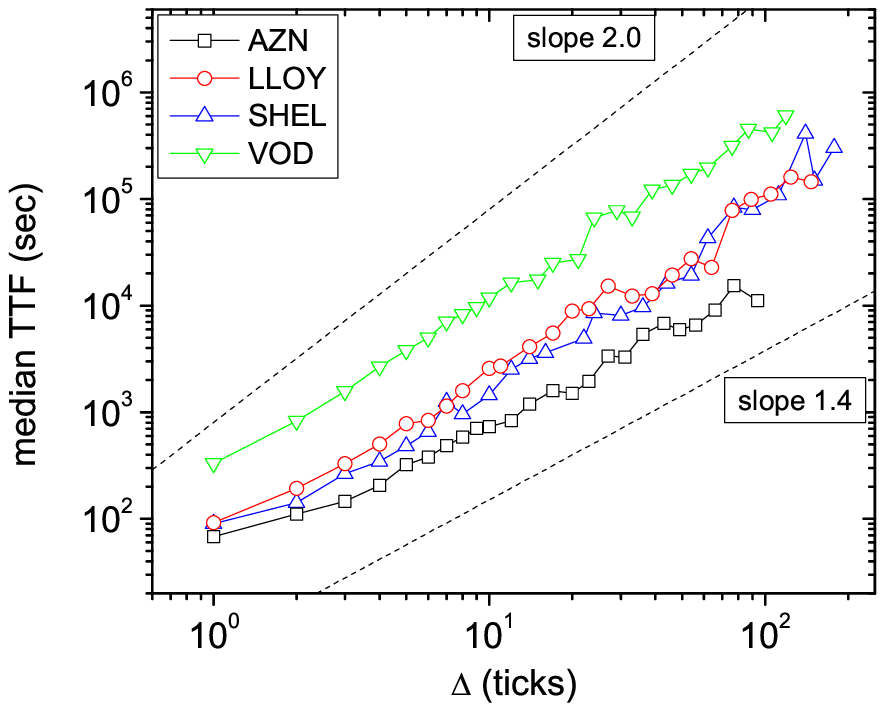}{\emph{Left}: The dependence 
of characteristic times on the entry depth $\Delta$ for GSK. Empty squares show 
time to fill, the curve for the surrogate first passage time assigned to filled 
orders is indistinguishable from this one. The dotted lines are $\med{\cdot} \propto \Delta^{1.4}$ and $\med{\cdot} 
\propto \Delta^{1.9}$. \emph{Right}: The dependence of median time to fill on 
the entry depth $\Delta$ for four stocks. The curves show a strong similarity 
and they are significantly better fitted by $\med{\TTF}\propto \Delta^{1.4}$ 
than by the naive ansatz $\med{\TTF}\propto \Delta^2$ from Brownian 
motion.}

\addfig{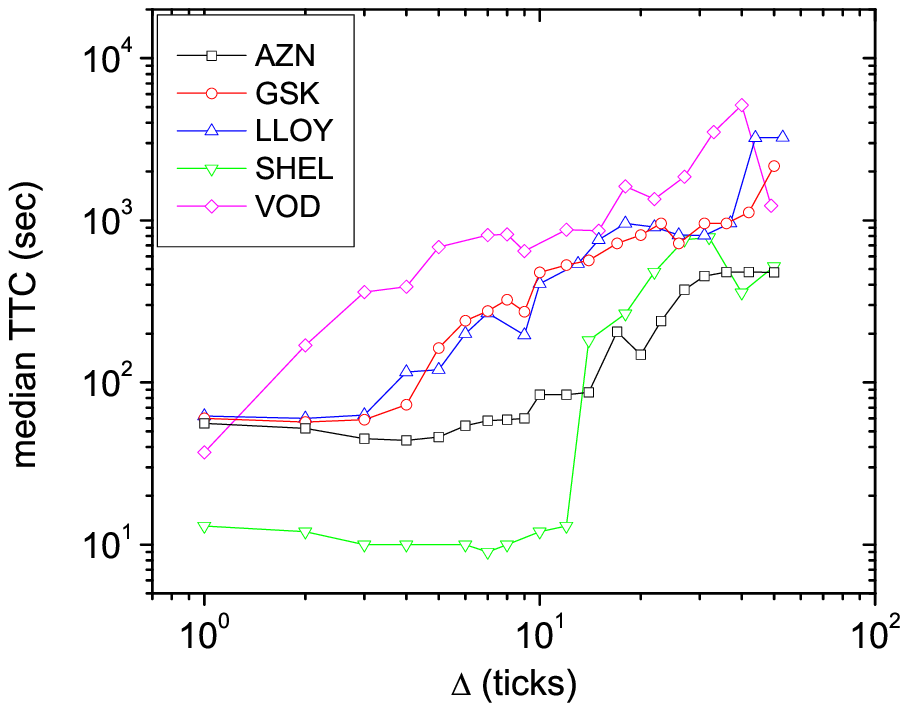}{The dependence of median time to cancel on 
the entry depth $\Delta$ of the limit order. The curves have an increasing 
tendency and they are qualitatively similar across stocks. However, they do not 
follow any obvious functional form.}

\section{A simple model of the characteristic times}

\label{sec:model}

The problem of the interplay between time to fill and time to cancel is an example for competing risks \cite{bernoulli.risk, hollifield.gains}. In this framework mutually exclusive events are considered in time \cite{bedford.competing, hollifield.gains}: in our case after its placement a limit order is either executed or canceled. Each of these events has its own probability distribution for the time when it will occur, but only the earliest one of the events is observed. In this section we present a simple joint model \footnote{Ref. \cite{bouchaud.relation} shows that similar arguments give a very good approximation for the average shape of the order book.} of limit order placement and cancellation that is of this type. We will see that the model gives predictions that can be tested against real data. Moreover, it also gives indications on the statistical properties of a quantity that is directly unobservable: the "lifetime" an agent is willing to wait for a limit order to be executed.

We make the following assumptions:
\begin{enumerate}
\item We consider one "representative agent" \cite{kirman.representative}. At time $t=0$ the agent places a single buy\footnote{Note that throughout the paper we use the language of buy orders, but analogous definitions can be given for sell orders. All measurements include both buy and sell orders.} limit order at a $\Delta > 0$ distance from the current best offer. (A generalization to $\Delta \leq 0$ is given in Section \ref{sec:moredelta}.) We treat all the other market participants on an aggregate level.
\item The agent is not willing to wait indefinitely for the 
order to be executed. Instead, at the time of placement the agent also decides 
about a cancellation (or more appropriately expiration) time $t'$ for the order. 
This is a value drawn randomly from the distribution ${P}_{\LT;\Delta}(t')$. We 
will call this function the \emph{lifetime distribution}. If the order is not 
executed until $t'$, then the order is canceled. The agent has no additional 
cancellation strategy. This assumption is very restrictive (cf. Ref. \cite{mike.empirical2}), but as Section \ref{sec:clock} will show, it does not affect our results significantly.
\item The market is very liquid and tick sizes are small. As a consequence,
\begin{enumerate}
\item before its execution, the effect of the agent's limit order on the evolution of the market price is negligible. This point neglects that traders reveal private information about their valuation of the stock by placing limit orders.
\item the interval between the time when the best bid reaches the order price and 
when the agent's order is executed is negligible. We also assume that such immediate execution is independent of the volume of the agent's order. A simple way to motivate that the volume present at a given price does not strongly affect execution times is to measure the typical ratio between time to fill and time to first fill as a function of the volume of the order. For at least $75\%$ of the orders of any volume this is close to $1$. The only exceptions can be very large orders with $\Delta = 1$. Here the price reaches the order quickly, but it takes about $20\%$ longer to execute it completely (see also Ref. \cite{lo.limitorder}). Moreover, for real limit orders the median time to fill does not depend too strongly on the volume of the order, except for very large volumes, see Fig. \ref{fig:Glifevol}.
\end{enumerate}
In our study we included SHEL and VOD which are known to have large tick/price ratios, so Assumption 3 would be invalid. Contrary to our expectations, we did not find any indication of anomalies like in other studies \cite{eisler.sizematters, eisler.liquidity, mike.empirical2,bouchaud.relation}, and the model proved useful for these stocks as well.
\end{enumerate}

\addtwofigs{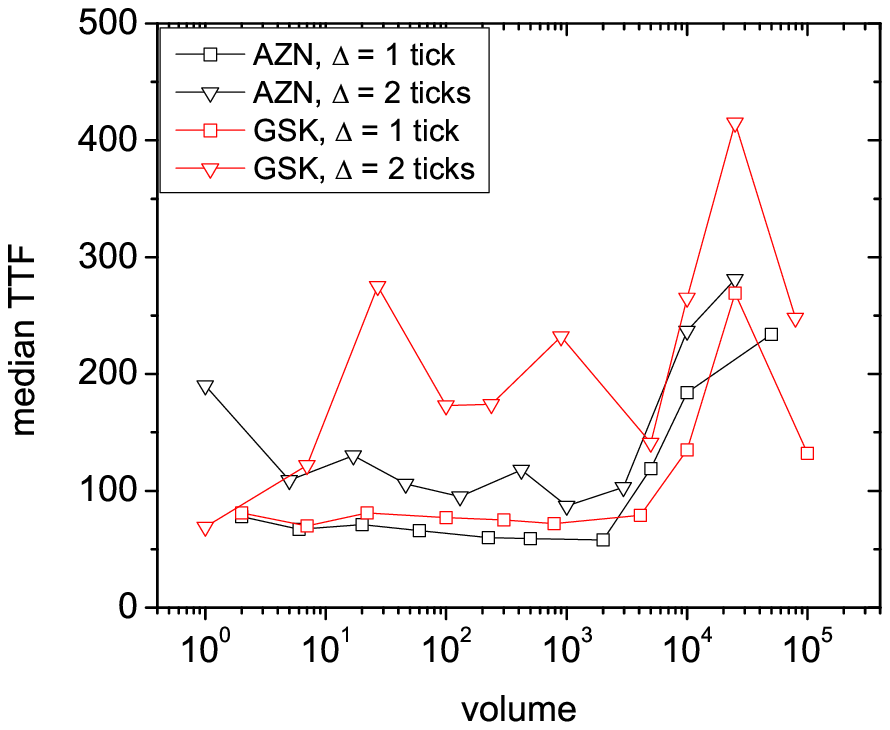}{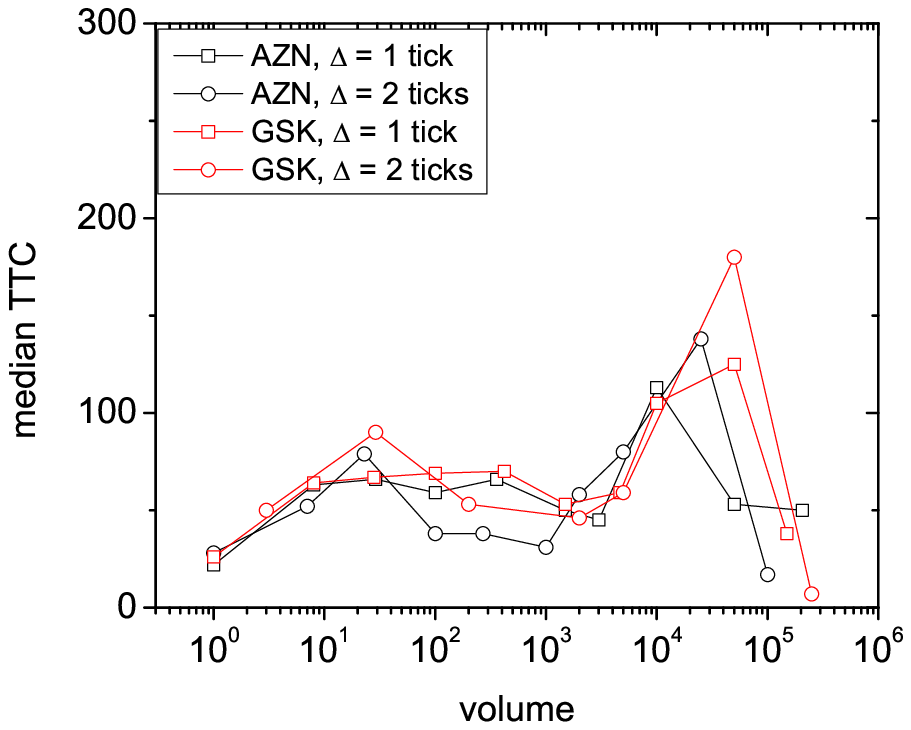}{\emph{Left}: median time to fill for 
AZN and GSK, $\Delta = 1$ and $2$ ticks, as a function of order volume. The value 
does not depend strongly on order size except for very large orders. 
\emph{Right}: median time to cancel for AZN and GSK, $\Delta = 1$ and $2$ ticks, as a function of order volume. The value does not depend strongly on order size. }

\begin{figure*}[ptb]
\centerline{\includegraphics[height=180pt]{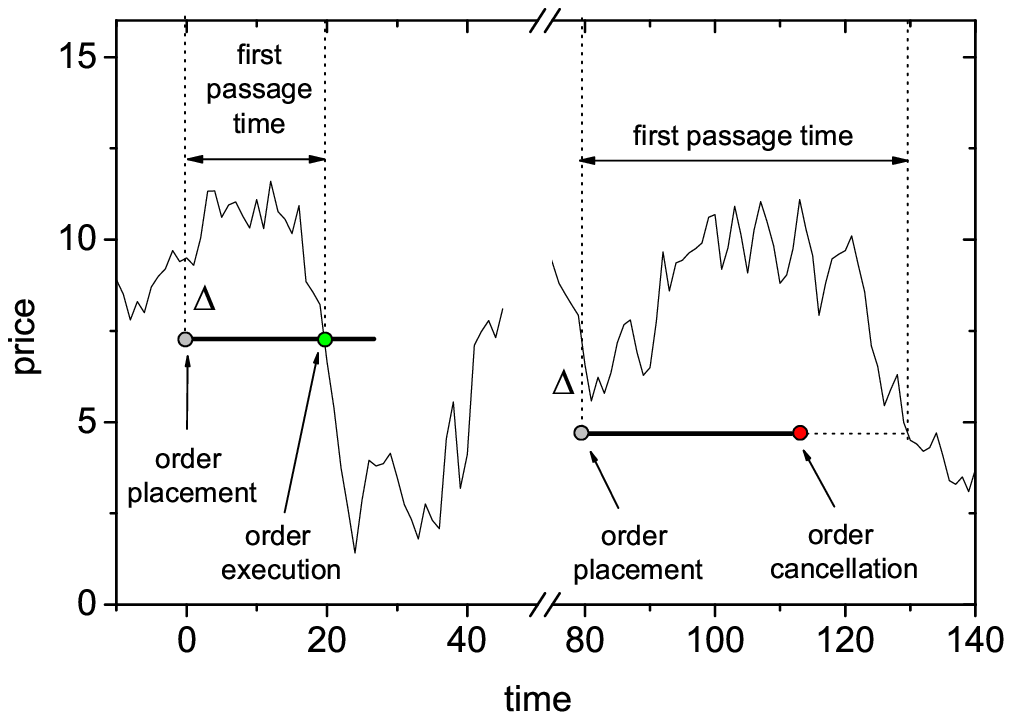}\includegraphics[height=90pt,trim=-10 55 80 70]{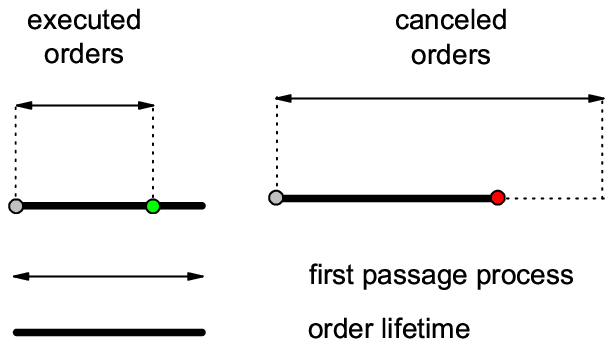}}
\caption{\emph{Left}: The scheme of the diffusive model for execution times (example with buy orders). Orders are indicated by thick horizontal lines. The order is placed at a distance $\Delta$ below the current 
best bid. At the time of its placement the order is assigned a lifetime 
(the length of the thick line). If the bid crosses the line, then the order is 
executed at the time of crossing. The time between order placement and the 
crossing is the first passage time of the bid price to a distance $\Delta$. If 
there is no crossing, the order is canceled at its cancellation time (the end of 
the thick line). \emph{Right}: Orders are executed when the first passage time 
is less, and canceled when larger than the intended lifetime.}
\label{fig:model}
\end{figure*}

Under our assumptions one can write a joint density function that describes both the price diffusion process and cancellations. The probability $P_\Delta (t,t')$ that the price reaches an order 
placed at a distance $\Delta>0$ from the current best offer at time $t$ (and then it can be executed immediately), and that the agent decides to cancel the order a 
time $t'$ can be written as a product of two independent 
distributions:
\begin{equation}
P_\Delta (t,t')=P_{\mathrm{FPT};\Delta}(t) P_{\LT;\Delta}(t').
\label{eq:pprod}
\end{equation}
For each limit order values of $t$ and $t'$ are drawn from $P$. The limit order is executed if $t<t'$ or it is canceled if $t>t'$. The two cases are illustrated in detail in Fig. \ref{fig:model}.

\section{The predictions of the model}
\label{sec:predictions}

Competing risk models are often estimated by the procedure introduced by Kaplan and Meier \cite{kaplan.meier}. This is a statistically consistent, non-parametric method to estimate the marginal distributions $\PFPT$ and $\PLT$ from $\PTTF$ and $\PTTC$ under the assumption that execution and cancellation are independent as we already assumed in writing Eq. \eqref{eq:pprod}. We will now calculate these estimates in another, but strictly equivalent analytical way.

Let us denote distribution functions as follows:
\begin{equation}
P_{X;\Delta}(>t)=\int_t^\infty  P_{X;\Delta}(\tau)d\tau,
\end{equation}
where $X$ can be any process introduced above (FPT, LT, TTF, TTFF, TTC). We will omit the lower index $\Delta$ for brevity. Let us first express the previously introduced quantities in terms of the joint 
probability $ P_\Delta (t,t')$ and via Eq. \eqref{eq:pprod}. For executed orders 
$t<t'$, thus the distribution of time to fill is given by
\begin{eqnarray} 
P_\mathrm{TTF}(t)=\frac{ P_\mathrm{FPT}(t) P_\LT(>t)}{\int_0^\infty 
P_\mathrm{FPT}(\tau) P_\LT(>\tau)d\tau}= \nonumber \\ \mathcal N[ P_\mathrm{FPT}(t) 
P_\LT(>t)].\label{eq:pf}
\end{eqnarray}
We introduced the operator $\mathcal N[\cdot]$, which normalizes a function to an integral of 
$1$. Symmetrically for time to cancel $t<t'$:
\begin{eqnarray} 
P_\mathrm{TTC}(t)=\frac{ P_\mathrm{FPT}(>t) P_\LT(t)}{\int_0^\infty P_\mathrm{FPT}(>\tau) P_\LT(\tau)d\tau}=\nonumber \\ \mathcal N[P_\mathrm{FPT}(>t) P_\LT(t)].
\label{eq:pc}
\end{eqnarray}

As \eqref{eq:pf} and \eqref{eq:pc} are two equations with only one unknown 
function, namely the lifetime distribution $ P_\LT(t)$, one can calculate that 
from, e.g., Eq. \eqref{eq:pf}, and then see if the solution is consistent with 
Eq. \eqref{eq:pc}. We can express from Eq. \eqref{eq:pf}, that
\begin{equation} 
P_\LT(>t) \propto \frac{ P_\mathrm{TTF}(t)}{P_\mathrm{FPT}(t)}
\label{eq:pbar}
\end{equation}
and thus
\begin{equation} 
P_\LT(t)=-\frac{d}{dt} P_\LT(>t)=-\mathcal N\left[\frac{d}{dt} \frac{P_\mathrm{TTF}(t)}{P_\mathrm{FPT}(t)}\right].
\label{eq:pcp1}
\end{equation}
It is also possible to estimate the same quantity directly from Eq. 
\eqref{eq:pc}:
\begin{equation} P_\LT(t)=\mathcal N\left[\frac{ 
P_\mathrm{TTC}(t)}{ P_\mathrm{FPT}(>t)}\right].
\label{eq:pcp2}
\end{equation}
Let us eliminate the lifetime distribution, and substitute the large $t$ asymptotic 
power law behavior of all probabilities. After simple calculations one finds 
that
\begin{equation}	\lambda_\mathrm{TTF}=\lambda_\mathrm{TTC}. \\	
\label{eq:lambdaright1}
\end{equation}
Then we substitute this result back into Eq. \eqref{eq:pbar} to find that the lifetime distribution also has to decay asymptotically as a power law:
\begin{equation}
P_\LT(t)\propto 
t^{-\lambda_\LT},	
\label{eq:pclock}
\end{equation}
with
\begin{equation}			
\lambda_\LT=\lambda_\mathrm{TTF}-\lambda_\mathrm{FPT}+1=\lambda_\mathrm{TTC}-\lambda_\mathrm{FPT}+1.	\label{eq:lambdaright2}
\end{equation}

Eq. \eqref{eq:lambdaright1} is in good agreement with the results of Section \ref{sec:ttfttc}, where $\lambda_\mathrm{TTF}=1.8-2.2$, and $\lambda_\mathrm{TTC}=1.9-2.4$. This is a clear improvement compared to Eq. \eqref{eq:lambdawrong}. The introduction of the simplest possible cancellation 
model gives a good prediction for the difference between the exponents describing the asymptotics of the 
first passage time and time to fill.

Moreover, one can now observe the hidden 
distribution of lifetimes. By substituting the typical values into Eq. 
\eqref{eq:lambdaright2}, one gets $\lambda_\LT \approx 1.6$. In comparison, a 
paper by Borland and Bouchaud \cite{Borland} describes a GARCH-like model 
obtained by introducing a distribution of traders' investment horizons and the 
model reproduces empirical values of volatility correlations for 
$\lambda_\LT=1.15$, which is not far from our estimate. More recently it has 
been shown \cite{Lillo06} that the limit order price probability distribution is 
consistent with the solution of an utility maximization problem in which the 
limit order lifetime is power law distributed with an exponent $\lambda_\LT 
\simeq 1.75$.  The origin of the power law distribution of limit order lifetimes is 
not clear. Unfortunately the data do not allow us to separate individual 
traders. Therefore we do not know whether such a result arises from the broad 
distribution of the time horizons of each trader, or simply a distribution of 
traders with different investment strategies. Based on an empirical 
investigation at the broker level, in Ref.~\cite{Lillo06} it is argued that 
heterogeneity of investors could be the determinant of the power law lifetime 
distribution. Notice, however, two points: (i) We are not speaking about how 
long the investors \emph{hold} the stock. Instead, $ P_\LT$ is the distribution 
of \emph{how long investors are willing to wait} for their limit orders to be 
executed and before they cancel or revise their offers. (ii) None of the limit 
orders we are discussing here are truly long-term. Even the orders with 
relatively long lifetime spend at most a few days in the book.

\section{An extension to $\Delta \leq 0$}\label{sec:moredelta}

So far we only considered orders with prices which were worse than the best offer at the time of their placement, i.e., $\Delta > 0$. However, this group only accounts for less than half of the 
actual limit orders. Measurements for $\Delta \leq 0$ orders give the surprising 
result that these execution times are described by statistics very similar to those for 
$\Delta > 0$. One example stock (GSK) is shown in Fig. 
\ref{fig:GttfGSKnegative}(left). The results of our fitting procedure performed with Eq. \eqref{eq:pttffit} are given in Table \ref{tab:ttfneg} for all five stocks.

\begin{table*}[tbp]
\centering
\begin{tabular}{c||ccc|ccc|ccc}\hline
stock & \multicolumn{3}{c|}{$\Delta = 0$} & \multicolumn{3}{c|}{$\Delta = -1$} & \multicolumn{3}{c}{$\Delta = -2$} \\ 
\cline{2-10} & $\lambda$ & $\lambda'$ & $T$ & $\lambda$ & $\lambda'$ & $T$ & $\lambda$ & $\lambda'$ & $T$  \\ \hline
AZN & $2.2$ & $0.6$ & $110$ & $2.2$ & $1.0$ & $230$ & $2.1$ & $1.2$ & $320$ \\
GSK & $2.2$ & $0.5$ & $110$ & $2.1$ & $1.1$ & $180$ & $2.2$ & $1.3$ & $410$ \\
LLOY & $2.2$ & $0.5$ & $120$ & $2.1$ & $1.0$ & $150$ & $2.0$ & $1.2$ & $220$ \\
SHEL & $2.2$ & $0.4$ & $110$ & $2.1$ & $1.0$ & $120$ & $2.1$ & $1.0$ & $140$ \\
VOD & $2.1$ & $0.5$ & $160$ & $2.0$ & $1.1$ & $130$ & -- & -- & -- \\ \hline
\end{tabular}	
\caption{Parameters of the fitting function \eqref{eq:pttffit} for the 
distribution of time to fill for the five stocks and $\Delta \leq 0$ ticks. All 
times are given in seconds. Data are missing when there were the statistics was 
inadequate for fitting. Typical standard errors for 
the quantities: $\pm 0.1$ for $\lambda_\TTF$, $\pm 0.1$ for $\lambda'_\TTF$, and 
$\pm 25\%$ for $T_\TTF$.}	
\label{tab:ttfneg}
\end{table*}

\begin{table*}[tbp]
\centering
\begin{tabular}{c||ccc|ccc|ccc}\hline
stock & \multicolumn{3}{c|}{$\Delta = 0$} & \multicolumn{3}{c|}{$\Delta = -1$} & \multicolumn{3}{c}{$\Delta = -2$} \\ 
\cline{2-10} & $\lambda$ & $\lambda'$ & $T$ & $\lambda$ & $\lambda'$ & $T$ & $\lambda$ & $\lambda'$ & $T$  \\ \hline
AZN & $2.3$ & $0.6$ & $130$ & $2.0$ & $0.7$ & $90$ & $1.9$ & $0.8$ & $72$ \\
GSK & $2.1$ & $0.6$ & $130$ & $1.9$ & $0.7$ & $90$ & $1.8$ & $0.8$ & $50$ \\
LLOY & $2.2$ & $0.5$ & $120$ & $1.9$ & $0.7$ & $70$ & $1.8$ & $0.9$ & $85$ \\
SHEL & $2.3$ & $1.0$ & $220$ & $2.2$ & $1.0$ & $130$ & $2.0$ & $1.0$ & $160$ \\
VOD & $2.0$ & $0.7$ & $200$ & $1.8$ & $0.8$ & $120$ & -- & -- & -- \\
\hline
\end{tabular}
\caption{Parameters of the fitting function \eqref{eq:pttcfit} for the distribution of time to cancel for the five stocks and $\Delta \leq 0$ ticks. All times are given in seconds. Data are missing when 
there were no orders at all, or the statistics was inadequate for fitting. 
Typical standard errors for the quantities: $\pm 0.1$ for $\lambda_\TTC$, 
$\pm 0.1$ for $\lambda'_\TTC$, and $\pm 25\%$ for $T_\TTC$.}	
\label{tab:ttcneg}
\end{table*}

According to our model, these orders should have 
been executed within a negligible time of their placement. While this is true for a 
number of them, certainly not for all. Let us assume that we are placing a new 
buy limit order. If our order has $\Delta = 0$, then it will be among the best 
offers at the time of its placement. If our order has $\Delta < 0$, then it 
becomes the single best offer in the book, and hence it will trade with 
certainty if the next event is a buy market order. Why can our order still take 
a long time before being executed? The answer is naturally that before our order 
is executed, a new buy limit order may enter the book. If this new order has 
$\Delta < 0$ (where $\Delta$ now has to be measured from our order),  it means that 
it has an even better price than our order and it 
will gain priority of execution. On the other hand, our order now effectively 
has $\Delta > 0$, and the original model can be applied.

In order to test such a hypothesis, we carried out the following calculation. For the sake of 
simplicity, we will consider the time to first fill instead of time to fill. 
Section \ref{sec:model} argued that for the majority of orders the difference 
between the two is negligible. From the time of its placement, we tracked every 
single at least partially filled $\Delta \leq 0$ order until the time it was 
first filled. We defined the reduced entry depth ($\Delta'$) and the reduced time 
to first fill ($\mathrm{TTFF}'$) for these orders as follows
\begin{enumerate}
\item For orders, where from their placement to their first fill there were no 
even more favorable orders both placed and then at least partially filled, 
$\Delta'=0$ and $\mathrm{TTFF}'=\mathrm{TTFF}$.
\item For orders where after their placement but before their first fill there was at least 
one new, more favorable order introduced with $\Delta_\mathrm{new}<0$ and then 
this new order was at least partially filled, we selected the first of such new 
orders placed after the original one and set $\Delta'=-\Delta_\mathrm{new}$. Thus, $\Delta'$ 
is the new position of the original order, after the new one was placed. 
$\mathrm{TTFF}'$ is defined as the time to first fill of our order measured from 
the placement of this new order.
\end{enumerate}

The typical distribution of $\mathrm{TTFF}'$ for different groups in $\Delta'$ is shown in Fig. 
\ref{fig:GttfGSKnegative}(right). For orders with $\Delta'=0$ this is -- except for here uninteresting
very short times -- well described by a stretched exponential 
distribution $P_\mathrm{TTFF'}(t)=\frac{1}{25}\exp\left[-\left(\frac{t}{6}\right)^{1/2}\right]$.
These are the orders, where there was no better offer made, and hence their 
execution times were purely determined by the incoming market orders. The 
distribution is very close to the distribution of the times between two 
consecutive transactions of the stock [see Fig. 
\ref{fig:GttfGSKnegative}(right)]. 

For orders with $\Delta'>0$, one recovers the 
results of the previous sections, and the distribution of reduced time to first 
fill asymptotically decays as a power law with a power close to $2.0$. Eqs. 
\eqref{eq:lambdaright1} and \eqref{eq:lambdaright2} are expected to be valid for 
orders with $\Delta < 0$ and $\Delta' > 0$ as well, given that we use them in 
terms of $\Delta'$ and $\mathrm{TTFF}'$. 

As a summary, time to first fill for 
orders with $\Delta \leq 0$ is a two-component process. If there is no better 
order placed before the first fill, then time to first fill is basically identical to 
the waiting time distribution between opposite market orders. If there is a 
better offer submitted, then the order effectively becomes $\Delta > 0$, and the 
diffusion approximation applies. As this latter process has a much fatter tail 
than the former one, long waiting times and the tail exponent of the joint process are again 
dominated by a first passage process.

\addtwofigs{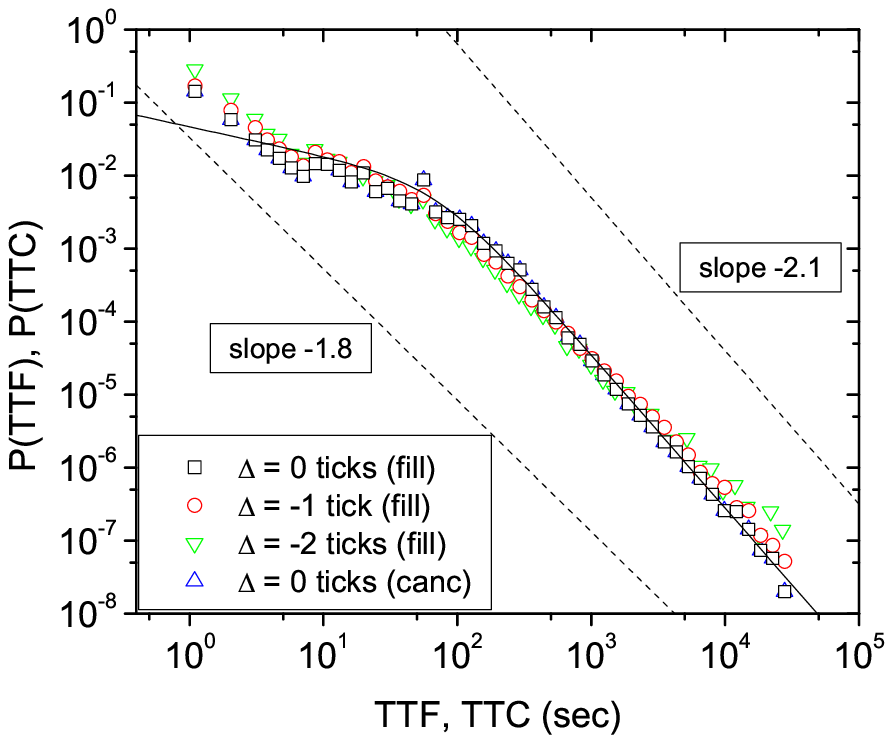}{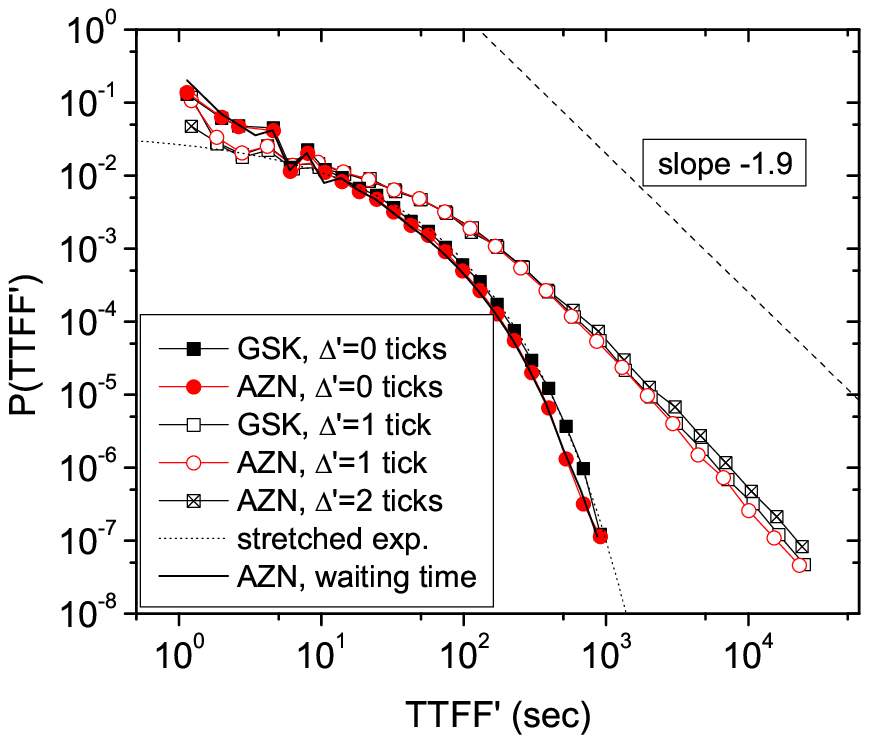}{\emph{Left}: Examples of 
the distribution of time to fill/cancel for $\Delta \leq 0$ buy limit orders of 
GSK. \emph{Right}: The distribution of reduced time to first fill 
($\mathrm{TTFF}'$) as a function of the reduced entry depth $\Delta'$. For 
orders where $\Delta'=0$, the tail of the distribution is well fitted by the 
stretched exponential $\frac{1}{25}\exp\left[-\left(\frac{t}{6}\right)^{1/2}\right]$. Where 
$\Delta'>0$, the distribution decays asymptotically as a power law with an 
exponent close to $-2.0$. The solid line is the distribution of waiting times 
between two consecutive trades of GSK.}

\section{Discussion}
\label{sec:conclusions}

\subsection{Lifetime distribution}
\label{sec:clock}
Before discussing the results let us analyze the most important simplifying assumption of our model, namely the independence of the lifetime of the order from the evolution of price. This would mean that traders decide about an expiry time of their limit orders 
at the time of their placement, and then do not cancel them earlier, which resembles the random cancellation process as introduced in Ref. \cite{zovko.farmer}. In order to see the relevance of our assumption
one should calculate the 
cross-correlation coefficient of first passage times and the lifetime process. However, as mentioned in Section \ref{sec:model} we are limited by 
the fact that the lifetime is hidden. It is not possible to calculate 
cross-correlations between time to fill and time to cancel either, because for the same order one cannot observe both variables. This issue is related to the identifiability problem of competing risks \cite{bedford.competing}. 

We suggest the following approach to resolve the above issue: Let us consider canceled orders only. There one can observe the values of the lifetime, because they were realized as an actual time to cancel. Moreover, our model assumed, that the order would have been executed at the first passage time (the time of the first transaction at the order's or a better price). Now it is possible to quantify cross-correlations between these two quantities, but one has to keep in mind three points. (Note that we will consider orders with $\Delta = 1$ to have the largest possible 
sample.)
\begin{enumerate}
\item For very short times the price dynamics is 
dominated by bid-ask bounce, and other non-diffusive processes \cite{Roll84}. 
Our model is not valid in this regime, because rapid order executions are not governed 
by a first passage process. Hence we discard all orders which were canceled 
within $L=4$ minutes of their placement.
\item In order to avoid problems arising from the possible non-existence of the moments of the distributions, we choose to evaluate Spearman's rank-correlation 
coefficient\footnote{This is defined by first, for both quantities separately, replacing each observation by its rank in the sample (i.e., assigning $1$ to the largest observation of first passage time, $2$ to the second largest, etc., and then repeating the procedure for lifetimes). Then the usual cross-correlation coefficient is calculated for the ranks \cite{lee.statistics}.} ($\rho$), instead of Pearson's correlation coefficient. The quantity $\rho$ has further favorable statistical properties, for example it is not very sensitive to extreme events.
\item As we can only consider canceled orders, we know that $\FPT>\LT$. 
This constraint alone, and regardless of the choice of correlation measure, will 
cause strong positive correlations between the two quantities. Even if $\FPT$ and $\LT$ are independent, the conditional joint distribution reads
\begin{eqnarray}
P(\FPT = t ,\LT = t'\vert \FPT > \LT) = \nonumber \\ \mathcal N[\Theta(t-t')\PFPT(t)\PLT(t')],
\end{eqnarray}
where $\Theta$ is the Heaviside step function. Due to our 
restricted observations this is clearly not a product of two independent 
densities.

Instead, a more convenient null hypothesis is to measure the correlations 
between $\FPT/\LT$ and $\LT$. $L=4$ min was chosen such that for $\Delta = 1$ the 
distribution of the first passage time is well described by the power 
law
\begin{eqnarray}
P_\FPT(t\vert t>L)\sim\frac{\lambda_\FPT-1}{L^{\lambda_\FPT-1}}t^{-\lambda_\FPT}. 
\label{eq:pttl}
\end{eqnarray}
If $\FPT$ and $\LT$ are independent, then
\begin{eqnarray} 
P\left(\FPT/\LT=x , \LT=t'\vert \FPT>\LT\right)=\nonumber \\ \mathcal N[\Theta(x-1)\PFPT(xt')\PLT(t')]= \nonumber \\ 
\mathcal N[\Theta(x-1)x^{-\lambda_\FPT}] \times \mathcal N[\PFPT(t')\PLT(t')].
\end{eqnarray}
Eq. \eqref{eq:pttl} was used for the second equality. The final result is a product form in functions of $x$ and of $t'$, which means that $\FPT/\LT$ is independent from $\LT$, given that we restrict ourselves to $\FPT>\LT$. Remember that the only assumption for this 
result is that first passage times are asymptotically power law distributed, 
which seems to hold very well in our data down to $L\approx 4$ min.
\end{enumerate}

We calculated Spearman's rank correlations between 
$\FPT/\LT$ and $\LT$ in our restricted sample for various stocks, this we will denote by $\rho_\mathrm{res}$. Results are 
summarized in Table \ref{tab:indep}. One 
finds negative correlation between the two quantities at all usual significance levels.\footnote{The error bars were estimated by the bootstrapping procedure suggested in Ref. \cite{schmid.bootstrapping} (for more details see Refs. therein).} This means that 
those limit orders that would have been executed later were canceled earlier, 
i.e., that traders update their decision on when to cancel a limit order by 
tracking the price path. This is in line with the results of Ref. 
\cite{mike.empirical2}. To prove that this value of $\rho$ truly comes from 
correlations, we generated surrogate datasets by randomizing the pairs $\FPT/\LT$ 
and $\LT$ while keeping the constraint $\FPT>\LT$. According to Table 
\ref{tab:indep} this completely destroys the correlations between $\FPT/\LT$ and 
$\LT$, $\rho_\mathrm{surr}=0$.

It is important to remember that this value of 
$\rho_\mathrm{res}$ is not the actual correlation coefficient between the first 
passage time and the lifetime process. To quantify the true value of 
cross-correlations, we introduce $\rho_\mathrm{true}$ which is Spearman's 
rank-correlation coefficient between $\LT$ and $\FPT$. While this cannot be 
measured directly, there is a procedure to estimate it from a known value of 
$\rho_\mathrm{res}$ based on Monte Carlo simulation. Let us assume that $\FPT$ 
and $\LT$ are adequately described by power law distributions with the known 
tail exponents. We model the cross-correlation between the two 
processes by copulas (see Ref. \cite{frees.understanding}). Morgenstern's copula reads
\begin{eqnarray}
P(>t, >t')=\PFPT(>t)\PLT(>t')\nonumber \\ \left\{1+3\rho_\mathrm{true}[1-\PFPT(>t)][1-\PLT(>t')]\right\},
\end{eqnarray}
with some $-1/3<\rho_\mathrm{true}<1/3$, while Frank's copula assumes
\begin{equation}
P(>t,>t')=\frac{1}{\alpha}\ln\left[1+\frac{(e^{\alpha \PFPT(>t)}-1)(e^{\alpha \PLT(>t')}-1)}{e^\alpha-1}\right],
\end{equation}
with some $-\infty < \alpha < \infty$. Here $P(>t, >t')=\int_t^\infty d\tau\int_{t'}^\infty d\tau' P(\tau, \tau')$ which is the joint distribution function. 

Monte Carlo measurements based on random pairs from these copulas suggest a nearly linear relationship between the true and the restricted correlation coefficients. With the substitution of the typical values of 
$\lambda_\FPT$ and $\lambda_\LT$ one finds that
\begin{equation}
\rho_\mathrm{true}=r\times\rho_\mathrm{res},
\end{equation}
where $r \approx 1.66$ for Morgenstern's and $r \approx 1.55$ for Frank's copula. 
The resulting estimates are given in Table \ref{tab:indep}. Naturally, the 
shuffled surrogate datasets yield $\rho_\mathrm{true}=\rho_\mathrm{res}=0$.

These calculations have shown that there is a strong negative correlation between the 
first passage time and the lifetime of an order in agreement with Ref. 
\cite{mike.empirical2} but contrary to our model assumption $2$ and 
Eq.~\eqref{eq:pprod}. So the key question is: How much does the presence of this 
correlation affect the predictions of our model? We performed a series of Monte 
Carlo simulations of the execution and cancellation processes by using the 
empirically observed value of tail exponents and cross correlations (Table 
\ref{tab:indep}). We found that for a fixed value of $\lambda_\FPT$ and 
$\lambda_\LT$ the introduction of such correlations increases the values of 
$\lambda_\TTF$ and $\lambda_\TTC$ by about $0.1$, which is comparable to the 
error bars of our estimates, and the power law behavior is well preserved. 
Moreover, the central part of our arguments, Eq. \eqref{eq:lambdaright1}, 
remains valid. Thus the presence of a dynamic cancellation 
strategy does not significantly affect the validity of our model.

\begin{table*}[tbp]
\centering
\begin{tabular}{c||c|c|c|c||c}\hline		
stock & $\rho_\mathrm{res}$ & $\rho_\mathrm{surr}$ & $\rho_\mathrm{true}$ 
(Morg.)& $\rho_\mathrm{true}$ (Frank)& number of points \\ \hline
AZN & $-0.12 \pm 0.02$ & $-0.001 \pm 0.001$ & $-0.19 \pm 0.03$ & $-0.18 \pm 0.03$ & $3277$ \\	
GSK & $-0.13 \pm 0.01$ & $0.000 \pm 0.001$ & $-0.21 \pm 0.02$ & $-0.20 \pm 0.02$ & $8573$ \\
LLOY & $-0.10 \pm 0.01$ & $0.000 \pm 0.001$ & $-0.16 \pm 0.02$ & $-0.15 \pm 0.02$ & $8201$ \\
SHEL & $-0.13 \pm 0.02$ & $-0.001 \pm 0.001$ & $-0.21 \pm 0.03$ & $-0.19 \pm 0.03$ & $2791$ \\
VOD & $-0.134 \pm 0.008$ & $0.000 \pm 0.001$ & $-0.22 \pm 0.01$ & $-0.21 \pm 0.01$ & $16392$ \\ \hline
\end{tabular}
\caption{Estimates of the correlation between first passage times and the lifetimes for orders with entry depth $\Delta = 1$ tick. We discarded orders executed or canceled within $L=4$ minutes 
of their placement. The errors represent the standard deviation estimated from $300$ bootstrap samples.}
\label{tab:indep}
\end{table*}

\subsection{Conclusions}

In this paper we focused on the tails of the distributions of characteristic times in the limit order book.
Our empirical observations, based on five highly liquid stocks on the London Stock Exchange, underline the importance of cancellations when comparing the first passage time to the time to execute an order. We found that the distributions follow asymptotically power laws for the first passage time, the time to (first) fill and time to cancel. The differences between the statistical properties of these characteristic times are informative of the interdependence of order executions and cancellations. These observations are quite robust and can be seen as "stylized facts" characterizing the order book. 

{We did not find significant difference between the behavior of buy and sell orders, in contrast with Refs.~\cite{lo.limitorder,cho.execution} for US markets, but in accord with Ref. \cite{hollifield.res2004} for the case of Ericsson stock traded at the Stockholm Stock Exchange. We are therefore not able to conclude whether the symmetric behavior we observe in the London Stock Exchange is common to most markets or specific to some of them or to certain time periods.}

In addition to the empirical findings summarized in Tables \ref{tab:fpt}, \ref{tab:ttf} and \ref{tab:ttc} we introduced a model, where order execution times are related to the first passage time of price, and orders are canceled randomly with lifetimes that are asymptotically power law distributed. This can be considered as the simplest possible model to take cancellations into account.  In this framework we showed that the characteristic exponents of the asymptotic power law behavior of the first passage time, the time to (first) fill and time to cancel are related to each other by simple rules which are in agreement with our empirical observations. These results are in contrast with another study (the NASDAQ data investigated in Ref. \cite{challet.pa2001}). Therefore further investigations are needed to clarify whether or not our findings are market specific.

{The observed heterogeneity of cancellation times may be driven by traders having different time horizons or by traders following different cancellation strategies in different market environments. Methods that can discriminate between these mechanisms represent a major objective for future research.}

\section*{Acknowledgments} We would like to thank two anonymous referees for useful comments and suggestions.
ZE is grateful to Jean-Philippe Bouchaud for discussions of the order book, to Michele Tumminello for advice on bootstrapping, and to Ingve Simonsen for help with the measurement of first 
passage times. The hospitality of l'Ecole de Physique des Houches and Capital 
Fund Management is also thankfully acknowledged. This work was supported by 
COST--STSM--P10--917 and OTKA T049238. FL and RNM acknowledge support from MIUR research project ``Dinamica di altissima frequenza nei mercati finanziari'' and NEST-DYSONET 12911 EU project.

\bibliographystyle{apsrev}
\bibliography{execution9}

\appendix
\section{Results in transaction time}
\label{app:ttime}

{The typical time between transactions strongly depends on market conditions and it is very far from strictly stationary. This fact, also closely related to volatility clustering, could influence the distribution of first passage times, time to fill and time to cancel. Many recent studies measure time in transactions in order to remove fluctuations in trading activity. In order to better understand the role of activity fluctuations, we repeated our calculations in transaction time, but we did not find any changes that affect the conclusions of our paper. Fig. \ref{fig:GfptGSK1tr} shows comparisons between real time and transaction time for the probability distributions of $\FPT$, $\TTF$ and $\TTC$ (for the stock GSK, $\Delta = 1$ tick). The short time regime is quite different, while for long times the fluctuations in trading activity are less relevant, and all the distributions remain power laws asymptotically. The changes in the values of the tail exponents are also small. The bottom right panel of Fig.  \ref{fig:GfptGSK1tr} compares $P_\mathrm{FPT}$, $P_\mathrm{TTF}$ and $P_\mathrm{TTF}$ in transaction time. Our arguments still hold, as $\lambda_\mathrm{FPT} < \lambda_\mathrm{TTF} \approx \lambda_\mathrm{TTC}$.}

\addfourfigs{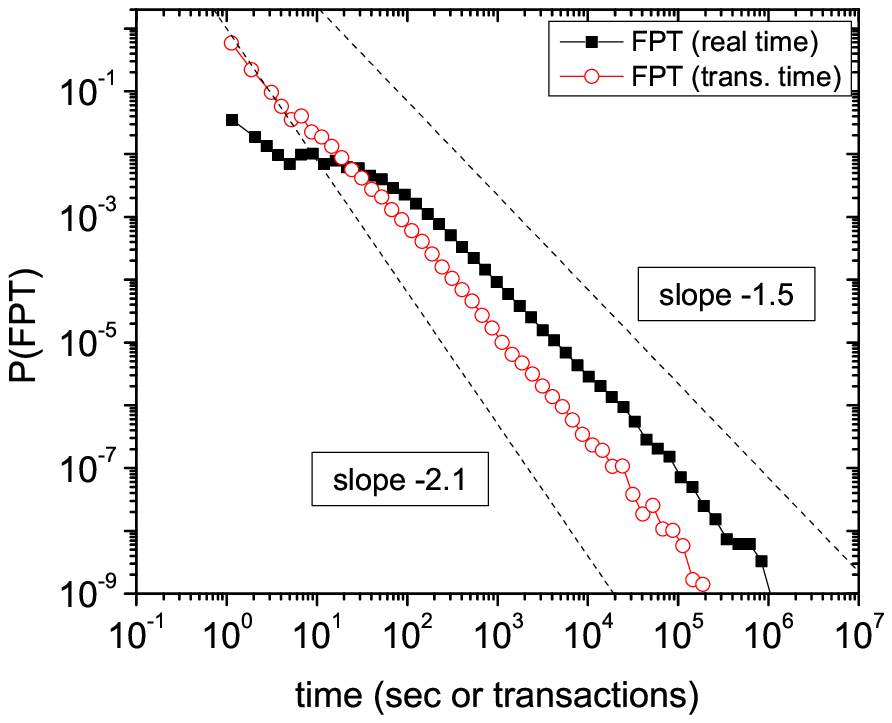}{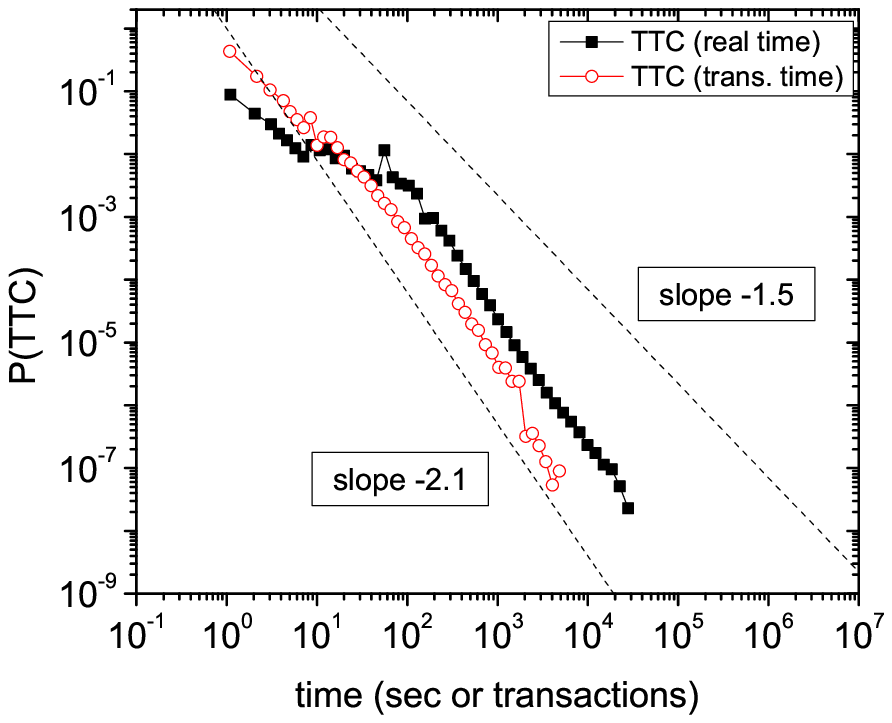}{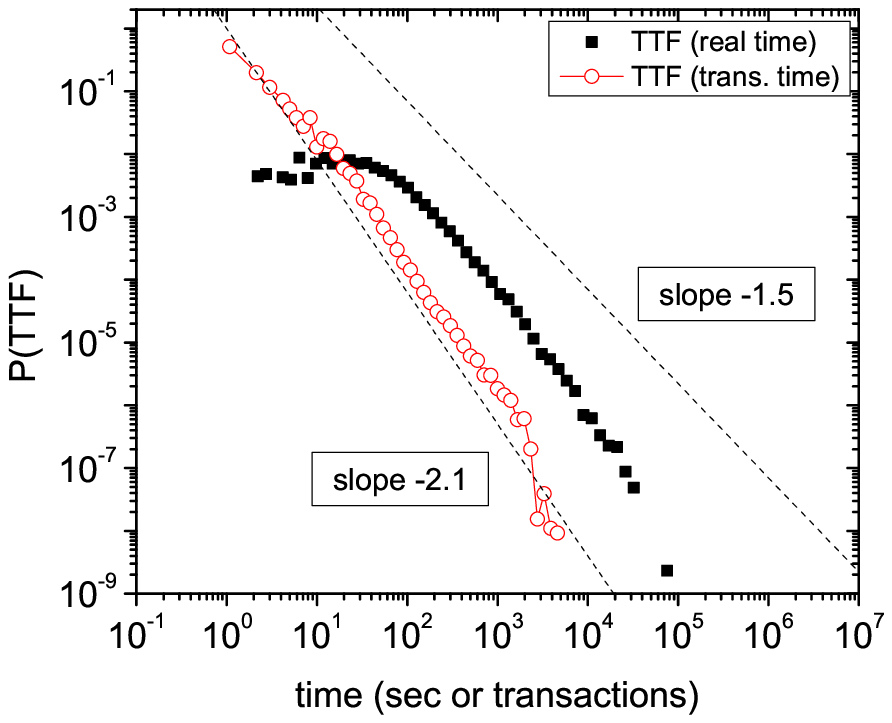}{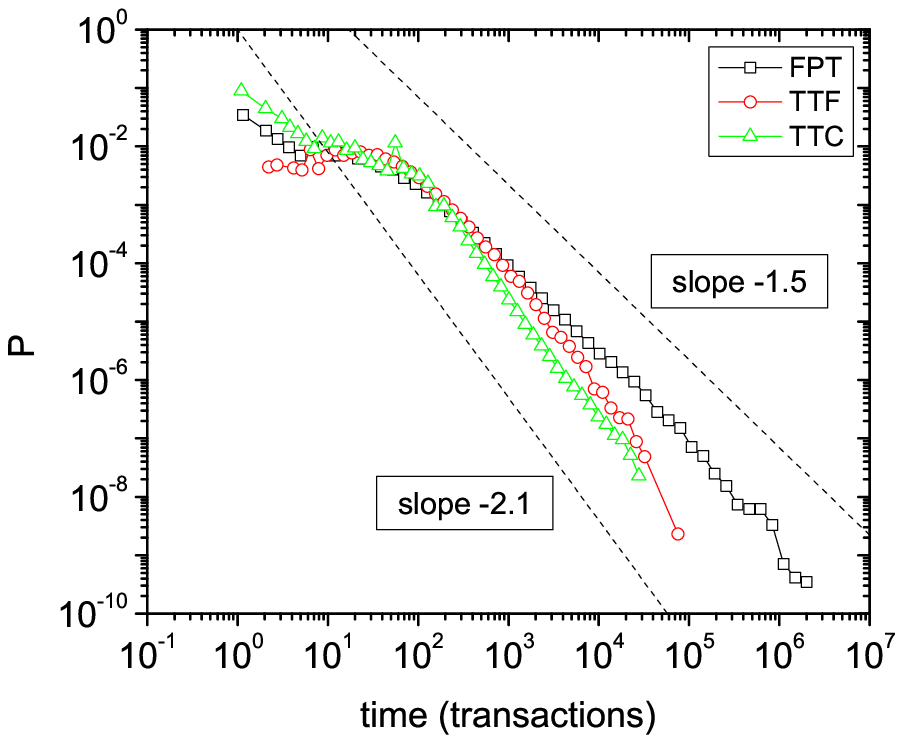}{Comparison of the distribution of characteristic times in real time and transaction time, all plots are for GSK, $\Delta = 1$ tick. As a reference two power law decays with exponents $1.5$ and $2.1$ are also given. \emph{Top left}: FPT, \emph{Top right}: TTF, \emph{Bottom left}: TTC, \emph{Bottom right}: all three quantities, only in transaction time.}

\section{Fitting functions for the distributions of time to fill and time to cancel}
\label{app:fit}
In this Appendix we present a critical discussion regarding our decision to fit the empirical distributions of FPT, TTF and TTC with the functional form presented in Eq.~\eqref{eq:pttffit}. In our preliminary investigations, we fitted the distribution of FPT, TTF and TTC with two different distributions. The first one was the one we consider throughout the paper, i.e.,
\begin{equation}     
P_Z(t) =
\frac{C't^{-\lambda}}{1+(t/T)^{-\lambda+
\lambda'}}.
\label{eqz}
\end{equation}
The second one was the generalized Gamma distribution
\begin{equation}
P_G(t)=\frac{\lambda|p|\kappa^\kappa(\lambda
t)^{p\kappa-1}\exp[-(\lambda t)^p\kappa]}{\Gamma(\kappa)},
\label{gammadist}
\end{equation}
which has been used in some of the existing studies on TTF (e.g. Ref. \cite{lo.limitorder}). Another common form, the Weibull distribution, is a special case of Eq.~\eqref{gammadist} for $\kappa=1$. Our empirical analysis shows that the Weibull distribution fits the
data poorly and it will not be considered in this Appendix.
For large values of $t$ the density of Eq.~\eqref{eqz} behaves as
\begin{equation}
P_Z(t)\sim\frac{1}{t^\lambda}~
\end{equation}
The asymptotic behavior of $P_G(t)$ depends on the sign of the
parameter $p$. If $p<0$ (as for the investigated data) it can be written as
\begin{equation}
P_G(t)\sim\frac{\exp(-c/t^{|p|})}{t^{1+|p|\kappa}}
\end{equation}
where $c$ is a constant. Thus the generalized Gamma distribution, similarly to Eq. \eqref{eqz},
is consistent with a power law tail, although it is modulated by an
exponential function which becomes less and less important as
$t\to\infty$. In order to estimate the optimal parameters of the
distributions we used a Maximum Likelihood Estimator (MLE). For illustrative purposes, here we consider the case of TTF for AZN and $\Delta=0$ but the results are similar for other stocks, other values of $\Delta$ and for both TTF and TTC.

\addtwofigs{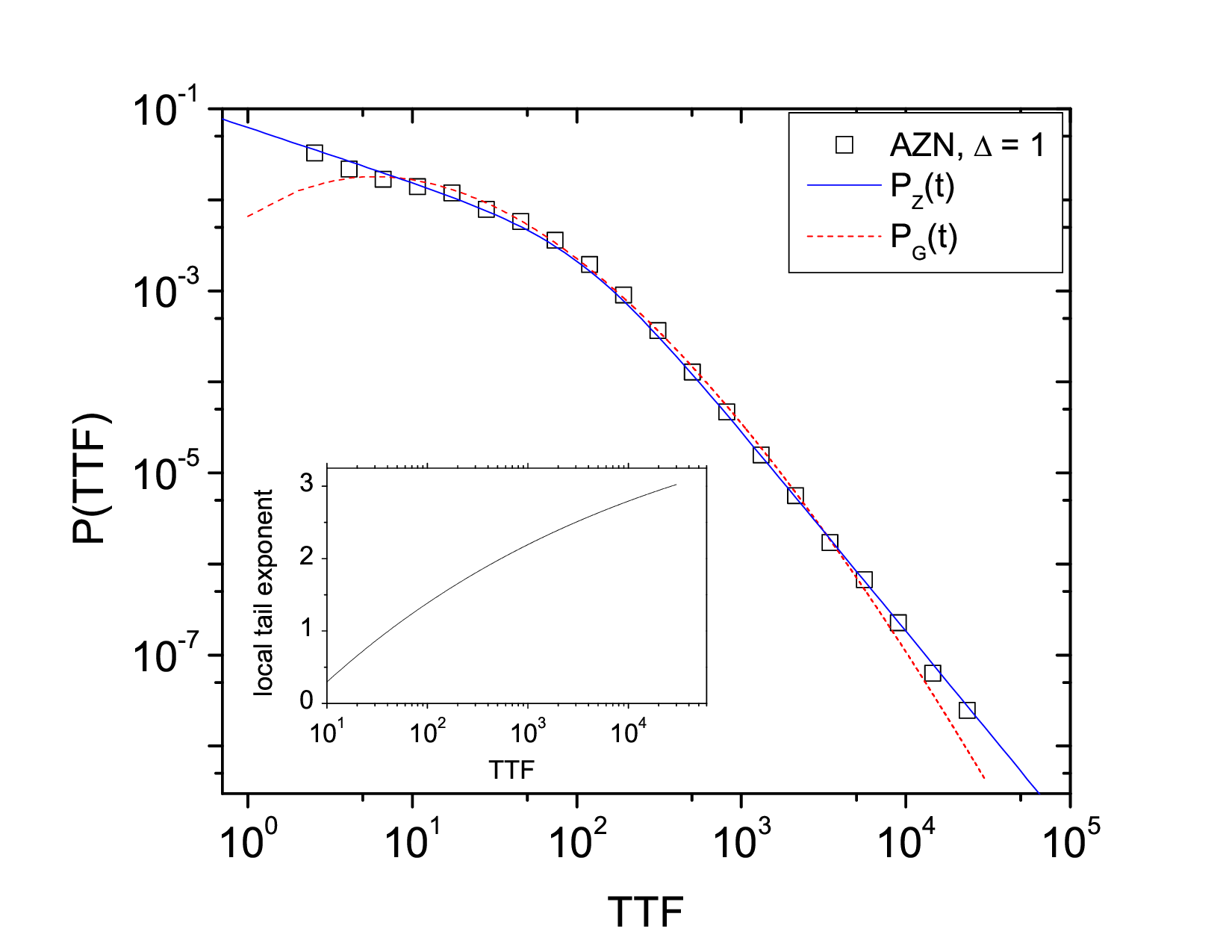}{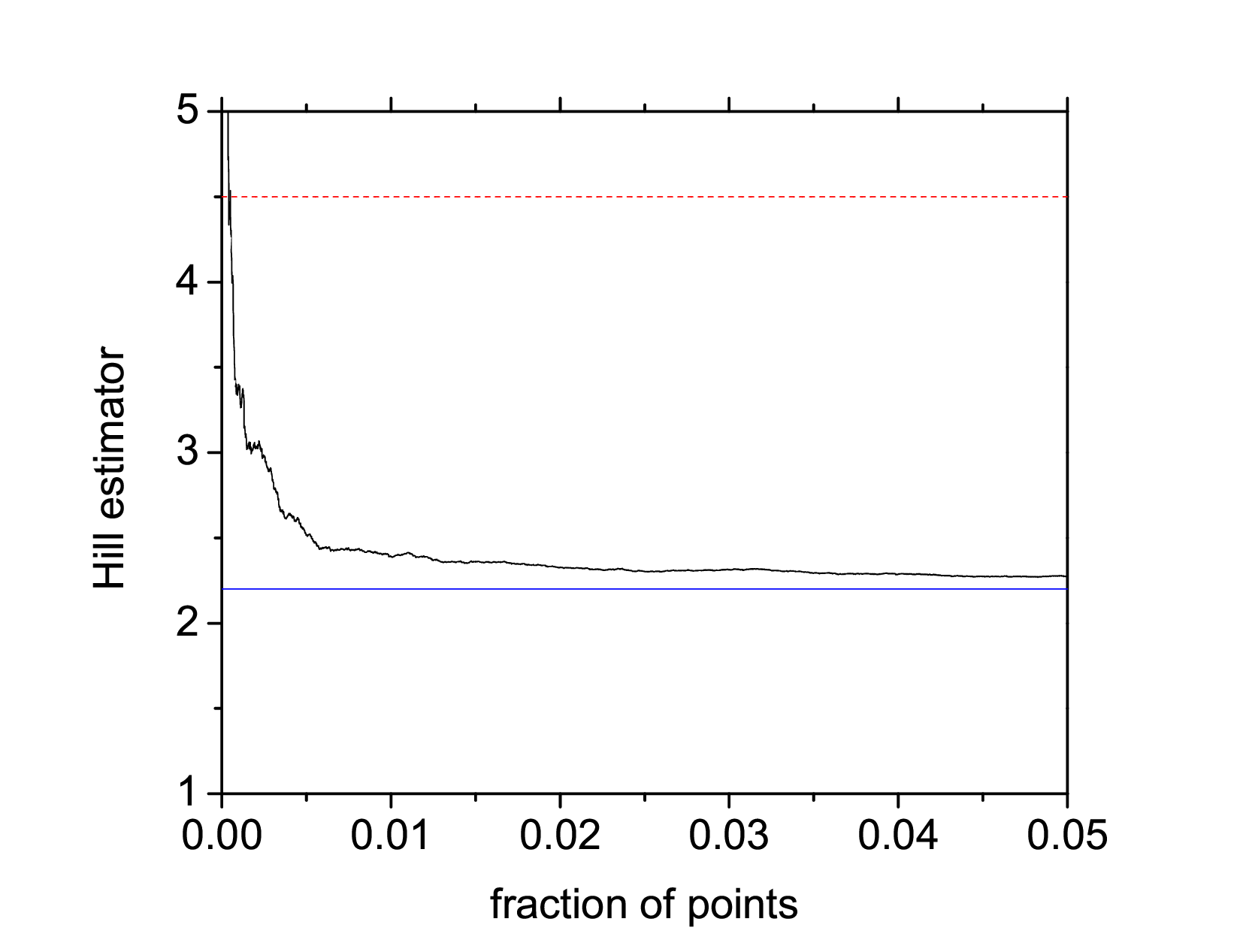}{\emph{Left}: Probability density function of the time to fill
of limit orders with $\Delta=0$ for AZN (boxes). We also show the Maximum
Likelihood best fit according to the generalized Gamma distribution
$P_G$ of Eq.~\ref{gammadist} (red dashed line) and to the functional
form $P_Z$ of Eq.~\ref{eqz} used in this paper (solid blue line). The
inset shows the local tail exponent of the generalized Gamma distribution ($d[\log P_G(t)]/d[\log t]$) as a function of $t$. \emph{Right}: Hill plot of the same data used in the left panel, showing the estimated tail exponent of the probability density function as a function of the fraction of points used in the estimation (black line). The dashed red line and the solid blue line are the
values of the exponent obtained from the fit by $P_G$ and
$P_Z$, respectively.}

Fig.~\ref{fig:Lapp}(left) shows the distribution of TTF
for AZN and $\Delta=0$ together with fits by Eqs. \eqref{eqz} and \eqref{gammadist}. Both $P_Z$ and
$P_G$  give a good fit both in the tail and in the body of
the distribution. One finds that
$P_G$ has a slightly larger likelihood ${\cal L}$ than $P_Z$. Since
the two distributions have the same number of parameters (degrees
of freedom) the likelihoods can be compared directly. However if one
computes the tail exponents of the distribution from the fitted
parameters one finds a puzzling result. The tail exponent
obtained from the generalized Gamma distribution fit is $4.5$, whereas
the tail exponent obtained from the $P_Z$ fit is $2.2$. Such a
difference in the exponent should be detectable in data. Still, Fig.~\ref{fig:Lapp}(left) shows that both distributions fit the tail reasonably well. The reason of this contradiction is shown in the inset of Fig.~\ref{fig:Lapp}(left). This plots the local tail exponent of the generalized Gamma distribution, given by $d[\log
P_G(t)]/d[\log t]$, as a function of $t$. The local exponent of the
generalized Gamma distribution converges extremely slowly to the
asymptotic value $4.5$ and in the range of the TTF from $10^3$ to
$10^4$ the local exponent is between $2$ and $3$, which is approximately
consistent with the values obtained from $P_Z$.

As we have repeatedly stated, in this paper we are interested in the
tail behavior of the distribution of the time to fill and time to
cancel. The analysis summarized
in Fig.~\ref{fig:Lapp}(left) shows that the parameters estimated from a fit to
a generalized Gamma distribution are not suitable to estimate the tail
exponent of the distribution, or at least not in the regime of TTF and TTC
values that can be explored within our dataset.  In other words, even if the
generalized Gamma distribution gives a (slightly) better fit in terms
of likelihood, it is hard to estimate the tail exponent from the fitted
parameters due to the slow convergence of the local exponent. On the
contrary, the parameters estimated from the fit with functional form of
Eq.~\eqref{eqz} give a better estimate of the tail exponent.
In order to support this claim, we estimate independently the tail exponent
by using the Hill estimator \cite{hill, embrechts}. In Fig.~\ref{fig:Lapp}(right) we show the Hill plot of the time to fill of
 AZN with $\Delta=0$. It is clear that the Hill estimator
converges to a value which is much closer to $2.2$ (as in the $P_Z$
distribution) than to $4.5$ (as in the $P_G$ distribution).

In conclusion our analysis shows that, although the generalized Gamma
distribution gives a slightly better \emph{overall} fit of time to fill and time to
cancel than our proposed form [Eq.~\eqref{eqz}], the parameters obtained
from the fit of $P_G$ suggest an unrealistic value of the tail exponent. On the contrary, our function $P_Z$ allows us to both fit the data reasonably well \emph{and} to obtain values of the tail exponent which are consistent with the Hill estimator. 

\end{document}